\documentclass{article}

\usepackage{amsthm}
\usepackage[english]{babel}
\usepackage{amsfonts}
\usepackage{amsmath}
\usepackage{amssymb}
\usepackage{graphicx}
\usepackage{graphics}
\usepackage{caption}
\usepackage{subcaption}
\usepackage{dcolumn}
\usepackage{bm}
\usepackage[margin=3cm]{geometry}

\begin{document}

\title{Ion-laser interactions: The most complete solution}

\author {H\'{e}ctor Moya-Cessa, Francisco Soto-Eguibar, Jos\'{e} M. Vargas-Mart\'{\i}nez\\
             \small Instituto Nacional de Astrof\'{\i}sica, \'Optica y Electr\'onica\\
             \small Calle Luis Enrique Erro No. 1\\
             \small Sta. Mar\'{\i}a Tonantzintla, Puebla. 72810 Mexico\\
             \\
             Ra\'{u}l Ju\'{a}rez-Amaro\\
             \small Universidad Tecnologica de la Mixteca\\
             \small Apartado postal 71\\
             \small Huajuapan de Leon, Oaxaca. 69000, Mexico\\
             \\
             Arturo Z\'{u}\~{n}iga-Segundo\\
             \small Departamento de Fisica, Escuela Superior de Fisica y Matematicas\\
             \small Edificio 9, Unidad Profesional Adolfo Lopez Mateos, IPN\\
             \small Mexico, DF. 07738 Mexico
              \\
              \\
              \\
              Published as: Physics Reports \textbf{513} (2012) 229-261}

\date{}
\maketitle

\begin{abstract}
Trapped ions are considered one of the best candidates to perform
quantum information processing. By interacting them with laser
beams they are, somehow, easy to manipulate, which makes them an
excellent choice for the production of nonclassical states of
their vibrational motion, the reconstruction of quasiprobability
distribution functions, the production of quantum gates, etc.
However, most of these effects have been produced in the so-called
low intensity regime, this is, when the Rabi frequency (laser
intensity) is much smaller than the trap frequency. Because of the
possibility to produce faster quantum gates in other regimes it is
of importance to study this system in a more complete manner,
which is the motivation for this contribution. We start
by studying the way ions are trapped in Paul traps in
and review the basic mechanisms of trapping. Then we
show how the problem may be completely solved for trapping states;
i.e., we find (exact) eigenstates of the full Hamiltonian. We show
how in the low intensity regime Jaynes-Cummings and
anti-Jaynes-Cummings interactions may be obtained, without using
the rotating wave approximation and analyze the medium and high intensity regimes were
dispersive Hamiltonians are produced. The traditional approach
(low intensity regime) is also studied and used for the generation of
non-classical states of the vibrational of the vibrational wavefunction. In
particular, we show how to add and subtract vibrational quanta to
an initial state, how to produce specific superpositions of number
states and how to generate NOON states for the two-dimensional
vibration of the ion. It is also shown how squeezing may be measured. The time dependent problem is studied  by using Lewis-Ermakov methods,
we give a solution to the problem when the time dependence of the
trap is considered and also analyze an specific (artificial) time
dependence that produces squeezing of the initial vibrational wave
function. A way to mimic the ion-laser interaction via classical optics is also introduced.
\end{abstract}

\section{Introduction}
The possibility to trap small clouds of particles, or inclusive to
trap individual atoms or ions, in an small region of space, was
opened with the invention of electromagnetic traps. These traps
allow to study isolated particles for long periods of time. The
Kingdon trap \cite{kingdon} is considered the
first type of trap developed, it consists of a metallic filament
surrounded by a metallic cylinder, and a direct current voltage
applied between them; the ions are attracted by the filament, but
its angular momentum makes them turn around in circular orbits,
with a low probability to crash against it. A dynamic version of
this trap can be obtained if also an alternate current voltage is
applied between the poles. However, this type of trap was not
widely used at that time, because it has short storage times and
because its potential is not harmonic. In 1936, Penning invented
another trap \cite{penning} in which the
action of magnetic fields together with electric fields make
possible the trapping of ions. The complete development of this
type of trap was reached when, in 1959, Wolfgang Paul designed an
electrodynamic trap (now called Paul trap) \cite{paul}.
In the Paul trap the idea  is that a charged particle can
not be confined in a region of space by constant electric fields,
instead an electric field oscillating at radio frequency, must be
applied. The Paul trap uses not only the focusing or defocusing
forces of the quadrupolar electric field acting on the ions, but
also takes advantage of the stability properties of the equations
of motion.
The ions trapped individually are very interesting, mainly because
they are simple systems to be studied. In particular, we take
advantage that  the ion motion in the Paul trap is approximately
harmonic, making  this system a simple one, allowing a better and
more direct comparison with theory. Individual ions of Ca+, Be+,
Ba+ and Mg+, can be storage, even for several days. The trapped
ions can be used to implement quantum gates, and a bunch of ions
arranged in a chain, is a promising tool to achieve a quantum
computer (each ion in the chain is a fundamental unit of
information or qubit) \cite{cirac}.
The trapping of individual ions also offers a lot of possibilities
in spectroscopy \cite{itano}, in the research of frequency
standards \cite{wineland,bollinger}, in the study of quantum
jumps \cite{powell}, the engineering of specific Hamiltonians \cite{vogeliii} and in the generation of nonclassical
vibrational states of the ion \cite{meekhof,moya94,moya99,Wall97,vogeliv,davidov,vogeli,vogelii}, to name some. To make the
ions more stable in the trap, increasing the time of confinement,
and also to avoid undesirable random motions, it is required that
the ion be in its vibrational ground state. This can be
accomplished by means of an adequate use of lasers; with the help
of these lasers, the internal energy levels of the trapped ion can
be coupled to their vibrational quantum states, in such a way,
that for a certain detuning, the coupling is equivalent to the
Jaynes-Cummings Hamiltonian \cite{JCM,paul2,Blockley,agarwal,JSM,cirac2}. On the other hand, the
beam that induces the coupling can be tuned to allow interactions
that generate simultaneous transitions of the internal and
vibrational states, either to lower vibrational energy levels
(while passing from the excited to the ground state) or to higher
vibrational energy levels (while passing from the ground to the
excited state); this type of coupling is called
anti-Jaynes-Cummings. Alternating successively Jaynes-Cummings and
anti-Jaynes-Cummings interactions, the trapped ion can be driven
to its vibrational ground state.
In this report, we study part of the physics of the trapped ions interacting with a laser
field. By using a set of time-dependent unitary transformations,
it is shown that this system is equivalent to the interaction
between a quantized field and a two level system with time
dependent parameters. The Hamiltonian is linearized in such a way
that it can be solved with methods that are found in the
literature, and that involve time-dependent parameters. The
linearization is free of approximations and assumptions on the
parameters of the system as are, for instance, the Lamb-Dicke
parameter, the time-dependency of the trap frequency  and the
detuning; thus, we can obtain the best solution for this kind of
systems. Also, we analyze a particular case of time-dependency of
the trap frequency.\\
Because we will assume an ion trapped in a Paul trap, in
Section 2, we review the basic mechanisms of it. In Section 3,
we show how exact eigenstates may be obtained for the ion laser system. In Section 4, we show how this interaction may be solved in different regimes. In Section 5, it is treated the standard approach to this interaction; i.e., the low intensity regime,
where by application of the rotating wave approximation, Jaynes-Cummings and anti-Jaynes-Cummings interactions may be
generated. By using this approach, we show how nonclassical states may be generated; in particular, we show how to add phonons to a vibrational state, how to filter specific superpositions of the motional wave function and how to generate NOON states in ions vibrating in two dimensions. We also propose here a
scheme that allows the measurement of squeezing. In section 6, we analyze the case of
an ion with a time-dependent frequency interacting with a laser
beam. By doing a series of unitary transformations, we linearize
the Hamiltonian of the system to an exact soluble form; this
linearization is also valid for any detuning and for any time
dependence of the trap.  In Section 7, we show how the ion-laser interaction may be modeled by evanescently coupling waveguides and Section 8 is left for conclusions.
\section{Paul trap}
\subsection{The quadrupolar potential of the trap}
As we already said, the Paul trap uses static and oscillating
electric potentials to confine charged particles. A charged
particle is linked to an axis if a linear restoring force acts
over it; i.e., if the force is
\begin{equation}\label{80010}
    \vec{F}=-c \vec{r},
\end{equation}
where $\vec{r}$ is the particle position and $c$ is a constant. In
other words, if the particle moves under the action of a parabolic
potential, that can be written in general form as
\begin{equation}\label{80020}
    \Phi(x,y,z)=A(\alpha x^2+\beta y^2 + \gamma z^2),
\end{equation}
where $A$ is another constant. The potential $\Phi$ must satisfy
the Laplace equation, which means that
\begin{equation}\label{80030}
    \nabla^2 \Phi=0,
\end{equation}
where $\nabla^2$ is the Laplacian operator. The Laplace equation
(\ref{80030}) imposes the condition
\begin{equation}\label{80040}
    \alpha+\beta+\gamma=0.
\end{equation}
To satisfy the above condition, we have several possibilities: \\
a) We make $\alpha=1$, $\beta=0$ and $\gamma=-1$, and this takes
us to the bidimensional potential
\begin{equation}\label{80050}
        \Phi=\frac{\Phi_0}{2r_0^2}(x^2-z^2).
\end{equation}
b) Another possibility is $\alpha=1$, $\beta=1$ and $\gamma=-2$,
and in this case we have, in cylindrical coordinates, the potential
\begin{equation}\label{80060}
    \Phi=\frac{\Phi_0}{r_0^2+2z_0^2}(r^2-2z^2),
\end{equation}
with $r_0^2=2z_0^2$. \\
\begin{figure}[h!]\label{Figure1}
  \centering
  \includegraphics [width=0.6\textwidth]{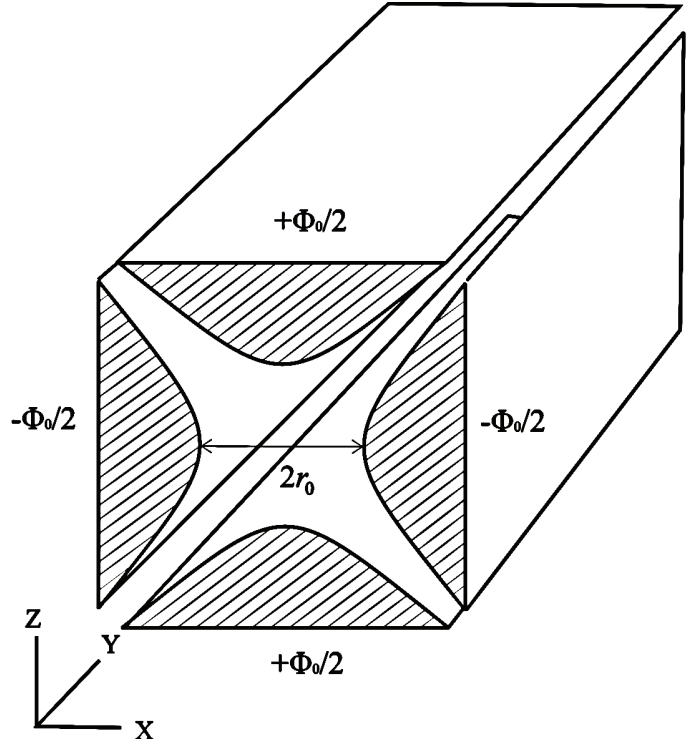}
  \caption{Electrode structure for the bidimensional configuration expressed by Equation (\ref{80050}).}
\end{figure}
Configuration a), Equation (\ref{80050}), is created by four hyperbolic electrodes linearly extended in the $z$ direction, as shown in Figure 1.
Configuration b), expression (\ref{80060}), is created by two electrodes with the form of an hyperboloid of revolution around $z$ axis. \\
The most used trap is the linear trap, as the one shown in Figure 1,
but with poles with circular transverse section instead of hyperbolic, because it is  easier to build.
This cylindric form does not correspond to some set of values of (\ref{80040}), but numerically it has been demonstrated that the potential produced by these electrodes near the axis of the trap is very similar to the one produced by the hyperbolic electrodes \cite{bonner}. \\
For the tridimensional case the magnitude of the field is given by
\begin{equation}\label{80070}
    E_x=\frac{\Phi_0}{r_0^2}x, \qquad  E_y=\frac{\Phi_0}{r_0^2}y, \qquad E_z=2\frac{\Phi_0}{r_0^2}z.
\end{equation}
Expressions (\ref{80060}) and (\ref{80070}) reveal that the
components $r$ and $z$ of the electric field are independent from
each other, and that they are linear functions of $r$ and $z$,
respectively. We also see that we have a harmonic oscillator
potential (parabolic and attractive) in the radial direction and a
parabolic repulsive potential in the $z$ direction. If a constant
voltage $\Phi_0$ is applied, and an ion is injected, the ion will
oscillate harmonically in the $x-y$ plane, but because the
opposite sign in the field $E_z$, its amplitude in the $z$
direction will grow exponentially. The particles will be out of
focus, and they will be lost by crashing against the electrodes.
Thus, the quadrupolar static potential, by itself, is not capable
to confine the particles in three dimensions; at most, with this
potential, we get unstable equilibrium. We will see next, how to
solve this problem.

\subsection{Oscillating potential of the trap}
To avoid the unstable behavior of the charged particles under a static potential, the trap must be modified. If an oscillatory electric field is applied, the particles can be confined. Because the periodic change of the sign of the electric force, we get focusing and defocusing in both directions of $r$ and $z$ alternatively with time. \\
If the applied voltage is given by a continuous voltage plus a voltage with a frequency $\Omega$, we have
\begin{equation}\label{80080}
    \Phi_0=U_0+V_0\cos\Omega t,
\end{equation}
and the potential in the axis of the trap is
\begin{equation}\label{80090}
    \Phi=\frac{U_0+V_0\cos\Omega t}{r_0^2+2z_0^2}(r^2-2z^2),
\end{equation}
where $r_0$ is the distance from the trap center to the electrode surface. \\
\begin{figure}[h!]\label{Figure2}
     \centering
     \includegraphics [width=0.9\textwidth]{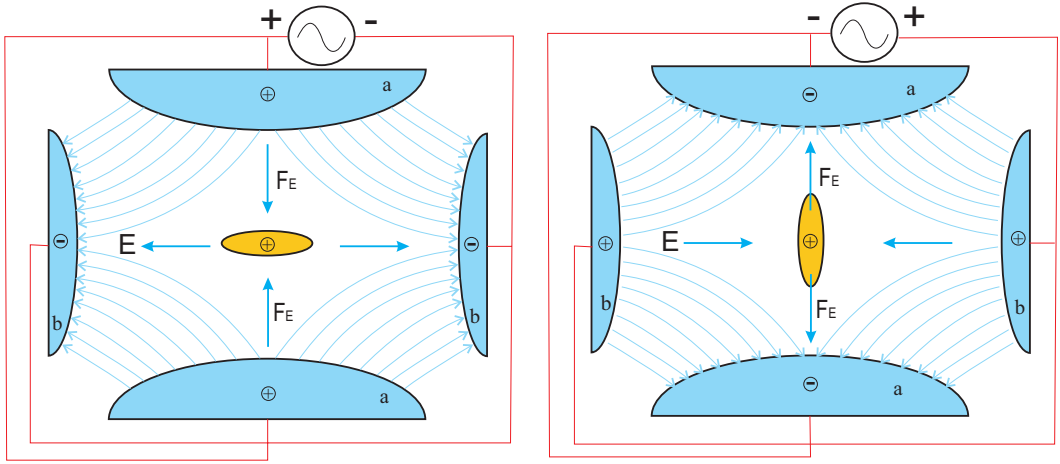}
      \caption{Scheme of a Paul trap to storage charged particles using oscillating electric fields generated by a quadrupole. The figure shows two states during an alternate current cycle.}
\end{figure}
In Figure 2, we show a transversal section of a Paul trap using an oscillating electric field.

\subsection{Motion in the Paul trap}
We will study now some details of the ion motion in a Paul trap. Let us consider the particular
case of just one ion, in three dimensions. If $m$ is the mass of the ion, and $e$ its charge, the equation of  motion is
\begin{equation}\label{80100}
    m\ddot{\vec{r}}(x,y,z)=q\vec{E}=-q\nabla\Phi.
\end{equation}
In order to analyze the trapping conditions, we write explicitly
each component,
\begin{equation}\label{80110}
      \ddot{x} = -\frac{2e}{mR^2}\left(U_0+V_0\cos\Omega t \right)x,
\end{equation}
\begin{equation}\label{80120}
      \ddot{y} = -\frac{2e}{mR^2}\left(U_0+V_0\cos\Omega t  \right)y,
\end{equation}
and
\begin{equation}\label{80130}
  \ddot{z} = \frac{2e}{mR^2}2\left( U_0+V_0\cos\Omega t   \right)z,
\end{equation}
where $R^2=r_0^2+2z_0^2$. \\
Making the substitution
\begin{eqnarray}\label{80140}
  a_r = \frac{8eU_0}{mR^2\Omega^2}=-\frac{a_z}{2}, \qquad  q_r = -\frac{4eV_0}{mR^2\Omega^2}=-\frac{q_z}{2},  \qquad
  \tau = \frac{\Omega t}{2},
\end{eqnarray}
equations (\ref{80110}) to  (\ref{80130}), take the
form of the Mathieu equation; i.e., take the form
\begin{equation}\label{80150}
    \frac{d^2x}{d\tau^2}+(a_r-2q_r \cos 2\tau)x=0,
\end{equation}
\begin{equation}\label{80160}
    \frac{d^2y}{d\tau^2}+(a_r-2q_r \cos 2\tau)y=0,
\end{equation}
and
\begin{equation}\label{80170}
    \frac{d^2z}{d\tau^2}+(a_z-2q_z \cos 2\tau)z=0,
\end{equation}
respectively. We can write the three equations as the following
one,
\begin{equation}\label{80180}
    \frac{d^2u_i}{d\tau^2}+(a_i-2q_i \cos 2\tau)u_i=0.
\end{equation}
The subindex $i=r,z$ corresponds to the quantities associated with
the axial and radial motions of the ion, respectively. The
quantities $u_i$ represent the displacement in the directions $r$
and $z$.

\subsection{Approximated solution to the Mathieu equation}
The Mathieu equation is a linear ordinary differential equation
with periodic coefficients. This equation can be solved using
Floquet's theorem \cite{bardroff}, which takes us to the general
solution
\begin{equation}\label{80190}
         u_i(\tau)=A_i e^{i\beta_i\tau} \phi(\tau)   +B_i e^{-i\beta_i\tau} \phi(-\tau),
\end{equation}
where $A_i$, $B_i$ and $\beta_i$ are constants determined by the
initial position, by the initial velocity of the ion, and by the trap parameters $a$ and $q$, and
\begin{equation}\label{80200}
    \phi(\tau)=\phi(\tau+\pi)=\sum_{n=-\infty}^{+\infty}C_n e^{2in\tau}
\end{equation}
 is a periodic function. \\
 The Mathieu equation has two types of solutions: \\
 1) Stable motion. When the characteristic exponent $\beta$ is real, the variable $u(\tau)$ is bounded, and in consequence the motion is stable. That means that the particle oscillates with bounded amplitudes and without crashing against the electrodes. These conditions allow to trap the ion. \\
 2) Unstable motion. When the characteristic exponent $\beta$ has an imaginary part, the function $u(\tau)$ has an exponential growing contribution. The amplitudes grow exponentially and the particles are lost when they crash against the electrodes.
\begin{figure}[h!]\label{Figure3}
    \centering
    \includegraphics [width=0.9\textwidth]{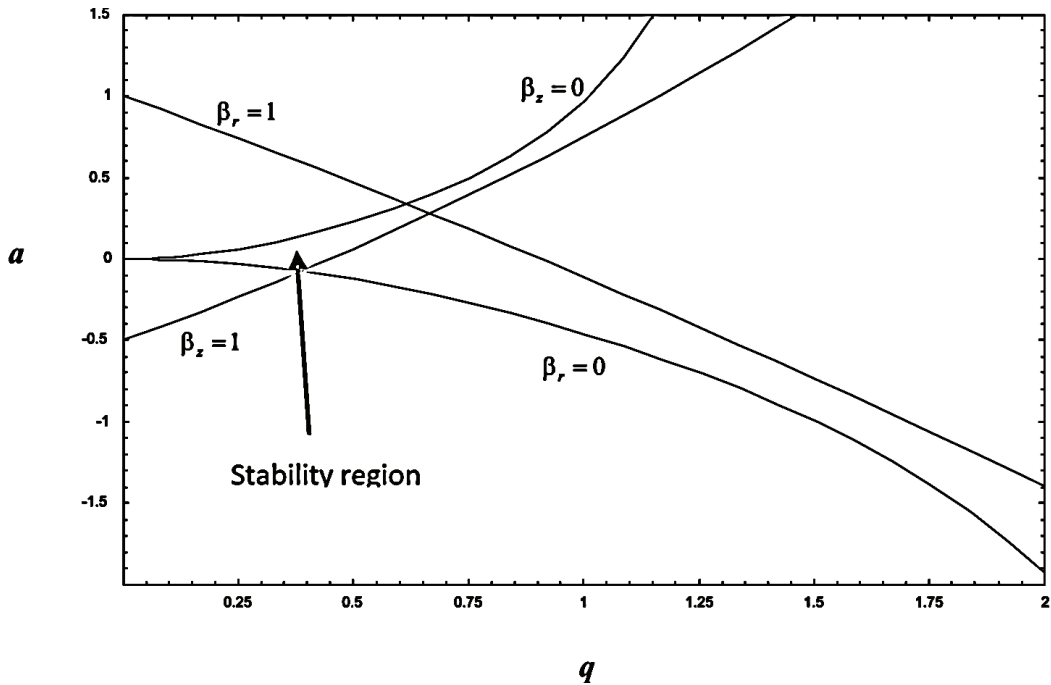}
    \caption{Stability region in the Paul trap.}
\end{figure}
The boundaries of the stability regions correspond to zero and integers values of $\beta_i$, and the first region of stability is surrounded by the four lines $\beta_r=0$ , $\beta_r=1$, $\beta_z=0$, and $\beta_z=1$, as is shown in Figure 3 \cite{Schleich}. \\
As $\beta_i$ is determined by $a$ and $q$, the Mathieu equation has stable solutions as a function of $a$ and $q$. Stability regions for the solutions of Equation (\ref{80180}) correspond to regions in the space of the parameters $a-q$, where there is an overlap of the stability regions in the axial and radial directions. \\
In the literature it is not possible to find analytic solutions
for Equation (\ref{80180}), but in most of the applications an
specification of the map of stability of the solutions is enough,
and it is not necessary a detailed functional dependence. However,
an approximated solution can be given in the stability region of
interest. To this end, we can write expression (\ref{80190}) as
\begin{equation}\label{80210}
    u_i(\tau)=A_i \sum_{n=-\infty}^{+\infty}C_{2n}^i \cos(2n+\beta_i)\tau
    + B_i \sum_{n=-\infty}^{+\infty}C_{2n}^i \sin(2n+\beta_i)\tau
\end{equation}
where, as we already said, $A_i$ and $B_i$ are determined by the
initial position $u_i(0)$ and initial speed $\dot{u}_i(0)$ of the
ion, respectively. The subindexes $i=r,z$ coincide with the
quantities associated with the radial and axial ion motion,
respectively.

The coefficients in the solution (\ref{80210}) are given by the
recurrence relations
\begin{equation}\label{80220}
       C_{2n+2}^i-D_{2n}^iC_{2n}^i+C_{2n-2}^i=0,
\end{equation}
with
\begin{equation}\label{80230}
       D_{2n}^i=\frac{a_i-(2n+\beta_i)^2}{q_i}.
\end{equation}
Once given $a_i$ and $q_i$, the quantities $C_{2n}^i$ and
$\beta_i$ can be calculated. If we define,
\begin{equation}\label{80240}
       G_{2n}^i=\frac{C_{2n}^i}{C_0^i}, \qquad A'_i=A_iC_0^i,  \qquad  B'_i=B_iC_0^i,
\end{equation}
and we make
\begin{equation}\label{80250}
        u_i(t)=u_i^s(t)+u_i^m(t),
\end{equation}
we get, from Equation (\ref{80210}),
\begin{equation}\label{80260}
     u_i^s(\tau)=A'_i \cos\omega_i t + B'_i \sin \omega_i t,
\end{equation}
and
\begin{equation}\begin{split}\label{80270}
       u_i^m(\tau)&=\sum_{n=1}^\infty    (A'_i \cos \omega_i t + B'_i \sin \omega_i t)
       (G_{2n}^i+G_{-2n}^i)\cos n \Omega t  \\
        & + (B'_i \cos \omega_i t - A'_i \sin \omega_i t)(G_{2n}^i-G_{-2n}^i ) \sin n \Omega t,
\end{split}\end{equation}
where $\omega_i=\beta_i \Omega/2$. \\
Analyzing Equations (\ref{80260}) and (\ref{80270}), it is
possible to realize that the ion motion has two components:
$u_i^s(t)$, a harmonic oscillation of frequency $\omega_i$, and,
$u_i^m(t)$, a superposition of several harmonics with a
fundamental frequency $\Omega$, and amplitudes modulated by the
frequency $\omega_i$. However, the proportion of the two
components, the values of $\omega_i$, the number of subcomponents
that contribute appreciably and their weights, depend strongly on
the values of $a_i$ and $q_i$, in such a way that they will change
in the stability region. All parameters are determined when the
values $\beta_i$ and $G_{2n}^i$ are given. Several values of
$\beta_i$ and $G_{2n}^i$, corresponding to some typical values de
$a_i$ and $q_i$, are listed in \cite{zhu}. In the table there, it
is possible to see that in the first region for $a \ll q \ll 1$,
we can assume that $G_2^i\cong G_{-2}^i$ and the rest of the
coefficients $G_{\pm 2n}^i$, $n>1$, can be ignored; thus,
Equations (\ref{80260}) and (\ref{80270}) can be rewritten as
\begin{equation}\label{80280}
        u_i^s(\tau)=u_{i0} \cos (\omega_i t + \delta _i),
\end{equation}
and
\begin{equation}\label{80290}
        u_i^m(\tau)=Cu_{i0} \cos \Omega t \cos (\omega_i t + \delta _i);
\end{equation}
or in other words,
\begin{equation}\label{80300}
        u_i(t)=u_{i0} \cos (\omega_i t + \delta _i) (1+C \cos \Omega t),
\end{equation}
with
\begin{equation}\label{80301}
      {u_{i0}}^2={A'_i}^2  + {B'_i}^2,
\end{equation}
\begin{equation}\label{80302}
      \delta = \arccos \left(  \frac{A'_i}{\sqrt{{A'_i}^2+{B'_i}^2}}  \right),
\end{equation}
and $C=2G_2^i$. \\
\begin{figure}[h!]\label{Figure4}
  \centering \includegraphics [width=0.85\textwidth] {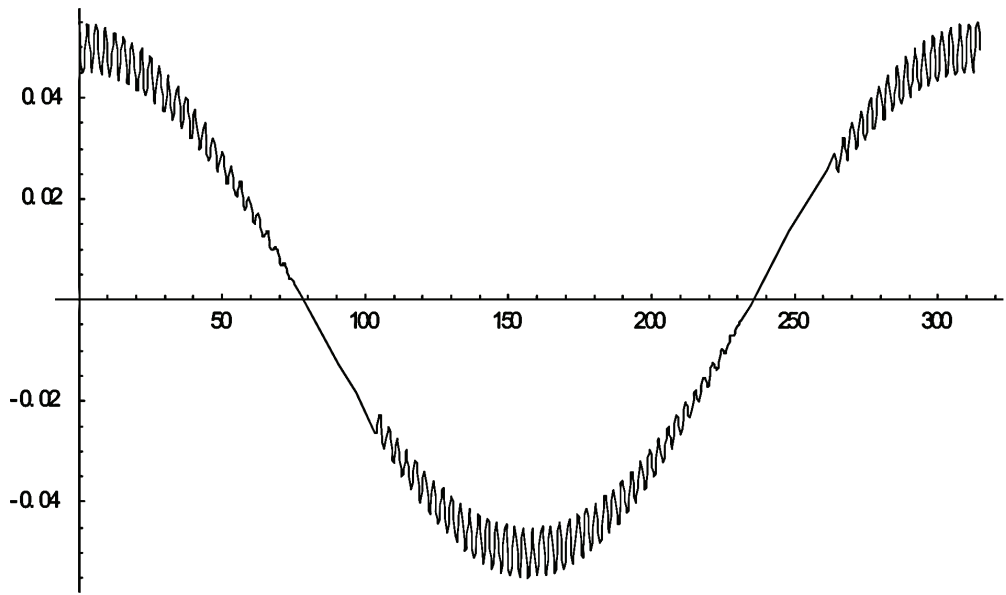}
  \caption{Micro motion and secular motion of a trapped ion with parameters $q=0.2, \beta=0.02$. The oscillations at high frequency are the micro motion and those at low frequency are the secular motion.}
\end{figure}
In Figure 4, we plot Equation (\ref{80300}). The ion motion is
composed by two types of oscillations: the harmonic oscillation
with frequency $\omega_i$, called secular motion, and the small contributions oscillating
at the frequency $\Omega$, called micromotion. Usually, the micromotion is ignored, but it
can be reduced using additional electrodes \cite{roos}. In this
way, the ion motion is controlled by Equation (\ref{80280}) and
behaves as it was confined in a harmonic pseudo-potential, that
for the radial part, has the form
\begin{equation}\label{80310}
    \Phi_r=\frac{m}{2}\left( \omega_x^2 x^2  +\omega_y^2 y^2   \right).
\end{equation}
Typically $U_0=0$, thus $a=0$ (Equation (\ref{80140})); in any
case, we are working in the region where $ a\sim 0$. Thus, the
frequencies $\omega_x$ and $\omega_y$ are degenerated, and
Equation (\ref{80310}) reduces to
\begin{equation}\label{80320}
    \Phi_r=\frac{m \omega_r^2}{2}\left(  x^2  + y^2   \right).
\end{equation}
To obtain an expression for $\omega_r$, we can use the
approximation \cite{dehmelt},
\begin{equation}\label{80330}
    \beta_r = \sqrt {a_r + \frac{q_r^2}{2}},
\end{equation}
with the definition $ \omega_r=\beta_r \Omega /2$, to get
\begin{equation}\label{80340}
    \omega_r=\frac{\Omega q_r}{2^{3/2}}=\frac{eV_0}{ \sqrt{2} m r_0^2 \Omega }.
\end{equation}
Experimentally, the typical ranges of operation are $V_0\approx300-800 \textrm{Volts}$, $\Omega/2\pi\approx16-18  \textrm{MHz}$, and $r_0\approx1.2 \textrm{mm}$, that gives a radial frequency $\omega_r\approx1.4-2.0 \textrm{MHz}$ for calcium ions $(^{40}\textrm{Ca}^+)$. \\
We can summarize this section, saying that under certain
conditions it is a good approximation to treat the ion motion as a
harmonic oscillator. This allows us in the following to apply algebraic techniques well known for this system
to the ion-laser interaction. In particular, invariant methods \cite{Lewis,Manueli,Manuelii} that permit the handling of
time dependent systems in a simple way.

\section{Ion-laser interaction in a trap with time-independent frequency}
Despite the relative simplicity of the ion-laser interaction, the full
theoretical treatment of its dynamics is a nontrivial problem, because this interaction is highly nonlinear. Even in the
simplest case, where only a single ion is in the trap, one is
usually forced to employ physically motivated approximations in
order to find a solution. A well-known example is the {\it
Lamb-Dicke} approximation, in which the ion is
considered to be confined within a region much smaller than the
laser wavelength. Many treatments also assume a {\it weak
coupling} approximation; i.e., a sufficiently weak laser-ion
coupling constant. Under these conditions, tuning the laser
frequency to integer multiples of the trap frequency results in
effective Hamiltonians of the Jaynes-Cummings type \cite{JCM,paul2,Blockley,agarwal,JSM,cirac2}, in
which the center-of-mass of the trapped ion plays the role of the field mode in cavity QED.
Recently, a new approach to this problem has been suggested, based on the application of a unitary
transformation $\hat{T}$ that linearizes the total ion-laser
Hamiltonian. Under this transformation the Hamiltonian becomes
{\it exactly} equivalent to the classic Jaynes-Cummings model
(JCM), including counter-rotating terms and an extra atomic
driving term. Remarkably, the ion-trap system is thus formally
equivalent to an atom interacting with a single-mode quantized
electromagnetic field. \\
This equivalence does not of course lead to an exact solution to
the ion-trap problem, since the complete JCM is also notoriously
unsolved. To our knowledge, eigenstates and eigenvalues for this
model have only ever been obtained either numerically  or
expressed in the form of series with no known closed expression. Still, some advantage can be taken of
regimes where the JCM is analytically, albeit approximately,
soluble, such as the well-known ``weak-coupling" limit where the
rotating-wave approximation can be made. Using the  already mentioned map
$\hat{T}^{\dagger}$ to translate this solution back into the
ion-trap scenario has led to the identification of a new soluble
regime for that system. This in turn has already
proven useful in the context of quantum computing. \\
In this section, we study the interaction of a laser with a trapped ion in a harmonic potential with constant frequency. We start with the Hamiltonian of the system, and we show that it is possible to find Jaynes-Cummings type transitions and anti-Jaynes-Cummings type transitions, depending on the different cases of resonance and laser intensities that induce the coupling between the ion internal states and the ion vibrational states. \\
We can write the Hamiltonian of the trapped ion as
\begin{equation}\label{3010}
     H=H_{\textrm{vib}} + H_{\textrm{at}} + H_{\textrm{int}},
\end{equation}
where  $H_{\textrm{vib}}$ is the ion's center of mass vibrational
energy, $H_{\textrm{at}}$ is the ion internal energy, and
$H_{\textrm{int}}$ is the interaction energy between the ion and
the laser. As we explained in the previous section, the
vibrational motion can be fairly approximated by a harmonic
oscillator. Internally, the ion will be modeled by a two level
system. In the interaction between the ion and the laser, we will
make the dipolar approximation, so we will write the interaction
energy as $-e\vec{r}\cdot\vec{E}$, where $-e\vec{r}$ is the
dipolar momentum of the ion and $\vec{E}$ is the electric field of
the laser, that will be considered a plane wave. Thus, we write
the Hamiltonian explicitly as
\begin{equation}\label{3020}
    H= \nu \hat{n} + \frac{ \omega_{21}}{2} \sigma_z + \lambda E_0 \left[ e^{i(kx-\omega t)} \sigma_+
    + e^{-i(kx-\omega t)} \sigma_- \right].
\end{equation}
The first term in the Hamiltonian is the ion vibrational energy;
in the ion vibrational energy, the operator
$\hat{n}=\hat{a}^\dagger \hat{a}$ is the number operator, and the
ladder operators $\hat{a}$ and $\hat{a}^\dagger$ are given by the
expressions
\begin{equation}\label{3030}
    \hat{a}=\sqrt{\frac{\nu}{2}} \hat{x} + i \frac{\hat{p}}{\sqrt{2 \nu}}
\end{equation}
and
\begin{equation}\label{3040}
    \hat{a}^{\dagger}=\sqrt{\frac{\nu}{2}} \hat{x} - i \frac{\hat{p}}{\sqrt{2 \nu}},
\end{equation}
where we have made the ion mass equal to 1. Also, for simplicity,
we have displaced the vibrational Hamiltonian by $ \nu/2$, the
vacuum energy, that in this case  is not important. \\
The second term in the Hamiltonian corresponds to the ion internal energy;
the matrices $\sigma_z$, $\sigma_+$, and $\sigma_-$ are the Pauli
matrices, and obey the commutation relations
\begin{equation}
   [\sigma_z,\sigma_{\pm}]=\pm 2\sigma_{\pm}, \qquad [\sigma_+,\sigma_-]=\sigma_z,
\label{pauli}
\end{equation}
and $\omega_{21}$ is the transition frequency between
the ground state and the excited state of the ion. \\
Finally, the third term, is the interaction energy between the ion and the
laser; in this last term, we have used again the rotating wave approximation.

\subsection{Exact eigenstates}
In this subsection we show that, under certain combinations of system
parameters, it is possible to obtain {\em exact} eigenstates for
the ion-trap Hamiltonian. Using the map $\hat{T}$ we also obtain
therefore exact eigenstates of the complete JCM. The set of states
we construct is by no means complete, and it is also unclear at
present how (or if) it may be extended. Nevertheless, we believe
that its existence may provide a clue to a deeper understanding of
both models. \\
Let us start by recalling the equivalence between the ion-trap
system and the JCM \cite{JCM,paul2,Blockley,agarwal,JSM,cirac2,dutra}. The Hamiltonian for the
ion-laser interaction can be written as
\begin{equation}\label{3050}
H_{\textrm{ion}}=\nu \hat{n}+\frac{\delta }{2}\sigma _{z}+ \Omega \left(
\sigma_{+} \hat{D}(i\eta)+\sigma_{-}
\hat{D}^{\dagger}(i\eta)\right),
\end{equation}
where $\hat{D}(i\alpha) = e^{i\alpha (\hat{a}+\hat{a}^{\dagger})}$ \cite{Glauber} is the
displacement operator, $\nu $ is the harmonic trapping frequency,
$\delta =\omega _{21}-\omega$ the laser-ion detuning,
$\Omega $ the (real) Rabi frequency of the ion-laser coupling and
\begin{equation}
    \eta = K \sqrt{\frac{1}{2m\nu}}
\end{equation}
is the so called Lamb-Dicke parameter, that  is a measure of the
amplitude of the oscillations of the ion with respect to the
wavelength of the laser field represented by its wave vector $K$. \\
On the other hand, the Jaynes-Cummings Hamiltonian with counter-rotating terms is given by
\begin{equation}\label{3060}
H_{JCM}=\omega \hat{n}+\frac{\omega _{0}}{2}\sigma _{z}+i\lambda
\left( \sigma _{+}+\sigma _{-}\right) \left( \hat{a} - \hat{a}^{\dagger
}\right).
\end{equation}
Although these two models appear to be physically and
mathematically quite distinct, they are in fact exactly equivalent.
The easiest way to see this is by rewriting Equation (\ref{3050}) in a
notation where operators acting on the internal ionic levels are
represented explicitly in terms of their matrix elements, as
\begin{equation}\label{3070}
H_{\textrm{ion}}=\left(
\begin{array}{cc}
\nu \hat{n}+\frac{\delta }{2} & \Omega \hat{D}\left( i\eta \right)
\\ \Omega \hat{D}^{\dagger }\left( i\eta \right) & \nu
\hat{n}-\frac{\delta }{2}
\end{array}
\right).
\end{equation}
Consider now the unitary operator
\begin{equation}\label{3075}
T=\frac{1}{\sqrt{2}}\left(
\begin{array}{ll}
\hat{D}^{\dag }\left( \beta\right) & \hat{D}\left( \beta\right) \\
-\hat{D}^{\dag }\left( \beta\right) & \hat{D}\left( \beta\right)
\end{array}
\right),
\end{equation}
where $\beta = i\eta/2$. It is possible to check after some
algebra that
\begin{equation}
{\mathcal H}_{\textrm{ion}}=TH_{\textrm{ion}}T^{\dagger }=\left(
\begin{array}{cc}
\nu \hat{n}+\Omega +\frac{\nu \eta ^{2}}{4} & \frac{\iota \eta \nu }{2}
\left( \hat{a} - \hat{a}^{\dag }\right) +\frac{\delta }{2} \\ \frac{\iota \eta
\nu }{2}\left( \hat{a} - \hat{a}^{\dag }\right) +\frac{\delta }{2} & \nu
\hat{n}-\Omega +\frac{\nu \eta ^{2}}{4}
\end{array}
\right).
\end{equation}
Returning to the usual notation, we obtain
\begin{equation}\label{3080}
{\mathcal H}_{\textrm{ion}}=\nu \hat{n}+\Omega \sigma _{z}+\frac{\iota \eta \nu }{2}
\left( \sigma _{+}+\sigma _{-}\right) \left( \hat{a}-\hat{a}^{\dag }\right) +\frac{
\delta }{2}\left( \sigma _{+}+\sigma _{-}\right) +\frac{\nu \eta^{2}}{4}.
\end{equation}
Comparing with Equation (\ref{3060}), it can be seen that this is
precisely the Jaynes-Cummings interaction, supplemented by two
additional terms: the first corresponds to an extra static
electric field interacting with the atomic dipole, and the second
is just a constant energy shift which can be
disregarded. In particular, a purely Jaynes-Cummings form is recovered when
$\delta =0$, corresponding to a resonant laser-ion interaction in
Equation (\ref {3050}). Of course, in Equation (\ref{3080}) the various
parameters in the Hamiltonian have different meanings than they do
in Equation (\ref{3050}): $\nu $ becomes the cavity field frequency
$\omega $, $2\Omega $ the atomic transition frequency $\omega
_{0}$, $\delta $ the coupling strength with the static field and
$\eta $ the ratio between the Jaynes-Cummings Rabi frequency
$2\lambda $ and the cavity frequency $\omega $. In what follows, we
shall refer to Equation (\ref{3080}) as the `Jaynes-Cummings picture'
of the ion-trap Hamiltonian, Equation (\ref{3050}). \\
This correspondence is very useful, since it enables one to map
interesting properties of each model onto their counterparts in
the other. For instance, it has been recently used to identify the
existence of ``super-revivals'' in the ion-laser interaction
\cite{dutra}, and to discover a means of realizing substantially
faster logic gates for quantum information processing in a linear
ion chain \cite{Jonathan}. In this paper, we will use it to map
eigenstates of one model into those of the other (it is clear
that, if $\left| \psi \right\rangle $ is an eigenstate of
$H_{\textrm{ion}}$, then $ T\left| \psi \right\rangle $ is a corresponding
one for ${\mathcal H}_{\textrm{ion}}$). In this regard it is important to
point out that, although $H_{\textrm{ion}}$ and ${\mathcal H}_{\textrm{ion}}$, both
describe systems consisting of a two-level atom interacting with a
bosonic mode, one should not identify each of these subsystems
with their counterparts after the transformation has been
applied. This is due to the fact that $T$ is an {\it entangling }
transformation: ion-trap states, where the ion's internal and
vibrational degrees of freedom have well-defined pure states, can
be mapped into entangled atom-cavity states in the corresponding
cavity QED system.

\subsubsection{Simple ansatz}
Let us return now to the ion-trap Hamiltonian, Equation (\ref{3050}).\
We will construct an {\it ansatz }which allows the determination
of exact eigenstates of this system, provided certain relations
are satisfied between the parameters $\Omega,\delta,$ and $\eta $. In
order to motivate our general solution, let us consider first the
possibility of finding such a state of the form
\begin{equation}\label{3090}
\left| \psi \right\rangle =\left| e\right\rangle \left( c_0\left|
0\right\rangle +c_1\left| 1\right\rangle \right) +\left|
g\right\rangle \left| \phi \right\rangle\equiv \left(
\begin{array}{c}
c_{0}\left| 0\right\rangle +c_{1}\left| 1\right\rangle \\ \left|
\phi \right\rangle
\end{array}
\right) ,
\end{equation}
where again we choose a notation where the ionic elements are
written out explicitly (e.g., $\left| e \right\rangle= $ $1
\choose 0 $). Let us now see whether the eigenvalue equation
\begin{equation}\label{eigveq}
H_{\textrm{ion}}\left| \psi \right\rangle =E\left| \psi \right\rangle,
\end{equation}
can be satisfied. Equation (\ref{3070}) shows that this requires
$\left| \phi \right\rangle $ to be of the form
\begin{equation}
  \left| \phi \right\rangle = D^{\dag }\left( i\eta \right) \left( d_{0}\left| 0\right\rangle
+d_{1}\left| 1\right\rangle \right) =
d_{0}\left|-i\eta\right\rangle +d_{1}\left|-i\eta; 1\right\rangle,
\end{equation}
where $\left|-i\eta\right\rangle$ is a coherent state and
$\left|\alpha ;k\right\rangle \equiv \hat{D}\left(\alpha \right)
\left| k\right\rangle $ is a displaced number state
\cite{oliveira}. We thus require
\begin{equation}\label{eigveq1}
H_{\textrm{ion}}\left| \psi \right\rangle =\left(
\begin{array}{c}
\left( c_{0}\frac{\delta}{2} +\Omega d_0 \right) \left|
0\right\rangle +\left(\Omega d_1 + c_{1}\left( \nu
+\frac{\delta}{2}\right) \right) \left| 1\right\rangle
\\ \left(c_{0}\Omega +d_0\left(\nu \hat{n}-\frac{\delta}{2}\right)\right)\left| -i\eta \right\rangle
+\left(c_{1}\Omega +d_1\left(\nu
\hat{n}-\frac{\delta}{2}\right)\right)\left| -i\eta
;1\right\rangle
\end{array}
\right).
\end{equation}
Now, using the well-known fact that $\hat{D}^{\dag
}\left(\alpha\right)\hat{a}\hat{D}\left(\alpha\right) = \hat{a}
+\alpha$ \cite{Glauber}, it is easy to show that displaced
number states satisfy the recursion relation
\begin{equation}\label{dispeq1}
\hat{n}\left| \alpha;k \right\rangle =(\left| \alpha \right|
^{2}+k)\left| \alpha;k \right\rangle +\alpha \sqrt{k+1}\left|
\alpha ;k+1\right\rangle +\alpha^* \sqrt{k}\left| \alpha
;k-1\right\rangle .
\end{equation}
Substituting then Equations (\ref{3090}), (\ref{eigveq1}) and
(\ref{dispeq1}) into Equation (\ref{eigveq}) gives the following
eigenstate conditions:
\begin{equation}\label{c0c1}
   d_1=0;\;\;c_{0}=\frac{\Omega }{\nu }; \;\;
   c_{1}=\frac{i\eta \nu }{\Omega };\;\;
   E= \nu+ \frac{\delta}{2},
\end{equation}
which hold however {\em only if} the parameters $\Omega ,\delta
,\eta $ satisfy the additional constraint
\begin{equation} \label{cond1a}
\left(\frac{\Omega}{\nu}\right)^2+ \eta^2 = 1 +
\frac{\delta}{\nu}.
\end{equation}
Under these conditions the state
\begin{equation}\label{eig1a}
\left| \psi_{\textrm{ion}}^+ \right\rangle = \left| e\right\rangle \left(
\frac{\Omega }{\nu }\left| 0\right\rangle +\frac{i\eta \nu}{\Omega }
 \left|1\right\rangle \right)
 +  \left| g\right\rangle \left| -i\eta \right\rangle.
\end{equation}
is an (unnormalized) eigenstate of $H_{\textrm{ion}}$ with eigenvalue $\nu
+ \delta/2$. Using operator $\hat{T}$ we can map this state into
an eigenstate of the generalized JCM model in Equation (\ref{3080})
\begin{equation}\label{eig1a10}
\left| \psi_{JCM}^+ \right\rangle = \hat{T}\left| \psi_{\textrm{ion}}^+
\right\rangle = \left| -\right\rangle \left( \frac{\Omega }{\nu
}\left|  -i\eta/2  \right\rangle +\frac{i\eta \nu }{\Omega }
 \left| -i\eta/2 ;1\right\rangle \right)
 +  \left| +\right\rangle \left| -i\eta/2 \right\rangle.
\end{equation}
where $\left| \pm\right\rangle = \left(\left| g\right\rangle
\pm\left| e\right\rangle\right)/\sqrt{2}$.\\
Condition (\ref{cond1a}) means that the {\em ansatz} in Equation
(\ref{3090}) does not always succeed, as only two of the three
parameters $\Omega ,\delta ,\eta $ can be chosen independently. In
addition, the domain of some of these parameters is not entirely
unrestricted; for instance, it is easily seen that no solution
exists when the laser is tuned such that $ \delta < -2\nu $.
Nevertheless, the existence of solutions satisfying
Equation (\ref{3090}) leads us naturally to seek for other solutions
using similar or slightly generalized ansatz. For
example, a second solution can be easily found if we note that the
Hamiltonian $ H_{\textrm{ion}}$ is invariant under the combined
transformations
\begin{equation}
 | e\rangle \leftrightarrow
| g\rangle, \qquad \delta \leftrightarrow -\delta, \qquad \eta \leftrightarrow -\eta  .
\end{equation}
 Applying this
symmetry transformation also to Equations (\ref{c0c1}) and (\ref
{cond1a}) we can see that, as long as we satisfy the condition
\begin{equation}\label{cond1b}
\left(\frac{\Omega}{\nu}\right)^2+ \eta^2 = 1 -\frac{\delta}{\nu},
\end{equation}
then
\begin{equation}\label{eig1b}
\left| \psi_{\textrm{ion}}^- \right\rangle = \left| e \right\rangle \left|
i\eta \right\rangle + \left| g \right\rangle \left( \frac{\Omega
}{\nu }\left| 0\right\rangle -\frac{i\eta \nu }{\Omega }\left|
1\right\rangle \right)
\end{equation}
is an eigenstate with eigenvalue $\nu -\delta/2 $.

Note that, unless $\delta =0$, conditions (\ref{cond1a}) and
(\ref{cond1b}) are {\it mutually exclusive}. (In the `ion-trap'
picture, this means the laser must be resonant with the ion; in
the 'JCM' picture, that no static field is applied; i.e., that the
Hamiltonian is just the full JCM). In other words, only in this
special case are $\left| \psi_{\textrm{ion}}^+ \right\rangle$ and $\left|
\psi_{\textrm{ion}}^- \right\rangle$ simultaneous (in fact, degenerate)
eigenstates of $H_{\textrm{ion}}$. The reason is that, in this case only,
the ion-trap Hamiltonian, Equation(\ref{3050}), has an extra symmetry:
it commutes with the operator $\sigma_x \exp(i\pi \hat{a}^{\dagger}\hat{a})$.
This parity-like observable has two eigenvalues $\pm1$, and so the
spectrum of $H_{\textrm{ion}}$ is two-fold degenerate (in the JCM picture,
the corresponding symmetry operator is $\sigma_z \exp(i\pi
\hat{a}^{\dagger}\hat{a})$) . It can be easily checked that neither $\left|
\psi_{\textrm{ion}}^+ \right\rangle$ nor $\left| \psi_{\textrm{ion}}^-
\right\rangle$ have this symmetry, but simple linear combinations
of them do.

\subsubsection{General ansatz}
One can easily generalize Equation (\ref{eig1a}) to obtain a more
general eigenstate for $H_{\textrm{ion}}$; this general state can be proposed as
\begin{equation}\label{33010}
    \left| \psi \right\rangle = \frac{\Omega }{\nu
    }\sum_{n=0}^{m+1}c_{n}\left| n\right\rangle \left| e\right\rangle
    +\sum_{n=0}^{m}d_{n}\left| -i\eta ,n\right\rangle \left|
     g\right\rangle.
\end{equation}
Substituting this proposal in the eigenvalue Equation (\ref{eigveq}), it can be shown that the corresponding eigenvalues are $E_m=(m+1)\nu +\frac{\delta}{2} $, that
\begin{equation}
    c_{n}=\left\{
    \begin{array}{c}
    \frac{1}{m+1-n}d_{n}; \qquad n=0,1,2,...,m \\
     i\eta \frac{\nu ^{2}}{\Omega ^{2}}\sqrt{m+1}d_{m};  \qquad  n=m+1
    \end{array}
     \right.
\end{equation}
and that the $d_{n}$ coefficients satisfy
\begin{equation}\label{33020}
    \left[
\begin{array}{ccccc}
\varepsilon _{0} & -i\eta  &  &  &  \\
i\eta   & \varepsilon _{1} & -i \sqrt{2}  \eta   &  &  \\
0 & i \sqrt{2} \eta  & \ddots  & \ddots  &  \\
&  & \ddots  & \varepsilon _{m-1} & -i \sqrt{m} \eta   \\
&  &  & i \sqrt{m} \eta  & \varepsilon _{m}
\end{array}
\right] \left[
\begin{array}{c}
d_{0} \\
\\
\vdots  \\
\\
d_{m}
\end{array}
\right] =\vec{0}
\end{equation}
where
\begin{equation}
    \varepsilon _n=1+m-n-\eta ^2+\frac{\delta }{\nu } -\frac{1}{1+m -n}\frac{\Omega ^2}{\nu ^2}.
\end{equation}
Equation (\ref{33020}) establish the recurrence relations
\begin{eqnarray}
    && i \sqrt{j} \eta  d_{j-1} + \varepsilon _j d_j - i\sqrt{j+1}\eta  d_{j+1}=0, \qquad   j=0,1,2,...,m-1\\ \nonumber
    &&  i \sqrt{m} \eta  d_{m-1} + \varepsilon _m d_m =0
\end{eqnarray}
that allow us to calculate all the $d$ coefficients of the eigenvector (\ref{33010}) in terms of $d_0$. The first ones are given by
\begin{eqnarray}
    d_1&=&-i \varepsilon _0 \frac{ d_0 }{\eta } \\ \nonumber
    d_2&=&\left(\eta ^2-\varepsilon _0 \varepsilon _1\right) \frac{d_0 }{\sqrt{2} \eta ^2} \\ \nonumber
    d_3&=&-i \left(2 \eta ^2 \varepsilon _0+\eta ^2 \varepsilon _2-\varepsilon _0 \varepsilon _1 \varepsilon _2\right) \frac{ d_0 }{\sqrt{6} \eta ^3} \\ \nonumber
    d_4&=&\left(3 \eta ^4-3 \eta ^2 \varepsilon _0 \varepsilon _1-2 \eta ^2 \varepsilon _0 \varepsilon _3-\eta ^2 \varepsilon _2 \varepsilon _3+\varepsilon _0 \varepsilon _1 \varepsilon _2 \varepsilon _3\right) \frac{d_0 }{2 \sqrt{6} \eta ^4}\\ \nonumber
\end{eqnarray}
Note that the vector of coefficients $\left(d_{0},...,d_{m}\right) $ is an eigenvector of this tridiagonal
matrix with zero eigenvalue. This is only possible if the determinant of the matrix in Equation (\ref{33020}) is zero, which imposes a constraint on $\Omega ,\delta ,\nu $.
These imposed conditions are the generalization of Equation (\ref{cond1a}). \\
It is easy to show that if $\Omega \ll \nu$, $\eta \ll 1$ and $\delta = - \nu$ all these conditions are satisfied (the matrix in Equation (\ref{33020}) becomes diagonal with a zero in the $m+1,m+1$ position), and then we obtain the \textit{exact} eigenvalues and the \textit{exact} eigenvectors without using the rotating wave approximation. \\
As we already mention, the Hamiltonian is symmetric under the combined transformations \\
 $\left\{ \left| e\right\rangle \leftrightarrow \left| g\right\rangle ;\text{ }\delta \leftrightarrow -\delta;\text{ }\eta \leftrightarrow -\eta \right\} $;
 this allow us to propose another set of eigenstates as
 \begin{equation}\label{33030}
    \left| \psi \right\rangle =\sum_{n=0}^{m}d_{n}\left| i\eta ,n\right\rangle \left|e\right\rangle
     +\frac{\Omega }{\nu}\sum_{n=0}^{m+1}c_{n}\left| n\right\rangle \left|g\right\rangle  ,
\end{equation}
where now
\begin{equation}
    c_{n}=\left\{
    \begin{array}{c}
    \frac{1}{m+1-n}d_{n}; \qquad n=0,1,2,...,m \\
     -i\eta \frac{\nu ^{2}}{\Omega ^{2}}\sqrt{m+1}d_{m}; \qquad n=m+1
    \end{array}
     \right.
\end{equation}
The $d$ coefficients satisfy Equation (\ref{33020}) but with
\begin{equation}
    \varepsilon _n=\frac{\delta }{\nu } +\frac{1}{1+m -n}\frac{\Omega ^2}{\nu ^2}-1-m+n+\eta ^2,
\end{equation}
and the corresponding eigenvalue are $E_m=(m+1)\nu - \frac{\delta}{2} $. \\
In this case all the constraints imposed by Equation (\ref{33020}) are satisfied if $\Omega \ll \nu$, $\eta \ll 1$ and $\delta =  \nu$, and the \textit{exact} eigenvalues and the \textit{exact} eigenvectors are again obtained without using the rotating wave approximation. \\

\subsection{Blue and red sidebands}
Processes where simultaneously  the internal excitation and the motional quantum numbers are increased
(decreased) are known as blue sideband excitations \cite{evers}. While, when simultaneously the internal excitation is excited (lowered) and  the motional quantum numbers are decreased (increased) are known as the red sideband. We discuss now some properties of the eigenstates (\ref{33010}) and (\ref{33030}) in order to show that, under certain conditions, namely, low intensity  ($ \Omega \ll \nu $) and Lamb-Dicke ($\eta  \ll 1$) regimes, eigenstates (\ref{33010}) correspond to the blue side band and the eigenstates (\ref{33030}) correspond to the red side band. Under these conditions is easy to prove that
\begin{equation}
    |\psi _0\rangle \approx i \eta \frac{\nu  }{\Omega }|1\rangle |e\rangle +|0\rangle |g\rangle,
\end{equation}
which is an eigenstate of the operator of the form  $\sigma _+\hat{a}^{\dagger }+\sigma _-\hat{a}$. Thus, when an up ion internal transition takes place ($|g\rangle \longrightarrow |e\rangle$) the vibrational motion acquires an extra phonon ($|0\rangle \longrightarrow |1\rangle$).
In general, we will have in this approximation
\begin{equation}
    |\psi _m\rangle \approx i \sqrt{1+m}\eta \frac{\nu  }{\Omega }d_m|m+1\rangle |e\rangle +\underset{n=0}{\overset{m}{\sum }}d_n|n\rangle |g\rangle
\end{equation}
showing that internal up transitions are accompanied by vibrational up transitions. \\
In the case of the eigenstates (\ref{33030}), we have the conditions $\Omega \ll \nu$, $\eta \ll 1$ and $\delta = \nu$. Under these conditions we can approximate the eigenstates as
\begin{equation}
    |\psi _m\rangle \approx  \underset{n=0}{\overset{m}{\sum }}d_n|n\rangle |e\rangle -i \sqrt{1+m}\eta \frac{\nu  }{\Omega }d_m|m\rangle |g\rangle,
\end{equation}
thus transitions to the lower internal ion states are associated with increase of vibrational quanta.

\subsection{The blue side band and the red side band by means of the intensity}
In the resonant case, processes that correspond to the blue side band and the red side band can also be obtained by means of the laser intensity. If we take $\delta = 0$, and consider the Lamb-Dicke limit, the displacement operator can be expanded in Taylor series and written as
\begin{equation}
\hat{D}(\alpha )=e^{\alpha  \hat{a}^{\dagger }-\alpha ^*\hat{a}}\approx 1+\alpha \hat{a}^{\dagger }-\alpha ^*\hat{a},
\end{equation}
such that the Hamiltonian (\ref{3070}) reads
\begin{equation}
    H=\nu  \hat{n}+\Omega  \sigma _x-\Omega \text{  }\eta  \sigma _y\left(\hat{a}+\hat{a}^{\dagger }\right),
\end{equation}
where we have used that $\sigma _x=\sigma _+ +\sigma _-$ and $\sigma _y=-i(\sigma _+ - \sigma _-)$.
By making now a rotation around the $Y$ axis (by means of the transformation $\exp(i\frac{\pi}{4}\sigma_y)$), and going to the interaction picture, we get the transformed Hamiltonian
\begin{equation}
    H_I=i \eta  \Omega  \left[ e^{-i t( \nu -2 \Omega )} \hat{a} \sigma _+- e^{i t (\nu -2 \Omega )} \sigma _- \hat{a}^{\dagger }- e^{-i t( \nu +2\text{  }\Omega )} \hat{a} \sigma _-+ e^{i t (\nu +2\Omega )} \sigma _+ \hat{a}^{\dagger } \right]
\end{equation}
If we take now $\nu =-2\Omega$, and we use the rotating wave approximation, the Hamiltonian reduces to
\begin{equation}
    H_I=-i \eta  \Omega  \left(\text{  }\hat{a} \sigma _- -\sigma _+ \hat{a}^{\dagger }\right)
\end{equation}
which clearly gives us the blue side band. \\
To get the red side band, we must take $\nu =2\Omega$ and the Hamiltonian we get is
\begin{equation}\label{33070}
    H_I=i \eta  \Omega  \left( \hat{a} \sigma _+-\sigma _- \hat{a}^{\dagger }\right)
\end{equation}
that clearly implies that when the ion goes from the ground internal state to the internal excited state, the vibrational motion losses one phonon and viceversa; i.e., the red sideband.

\section{Solution in different regimes: dispersive Hamiltonians}
The ion-laser interaction may be easily solved in the low intensity regime
\cite{wine2,wine3,ion,Buzek,tomb}, but besides the
condition that the laser intensity is much lower than the
vibrational frequency, we set the condition that the detuning
between the laser and the atomic transition frequency is an
integer multiple  of the vibrational frequency. Then some
questions arise: Is it possible not to consider integer multiples
of the vibrational frequency?, Is it possible to solve for high and
middle intensities?

Indeed, we have shown in Section 3 that  it is possible to find solutions for any set of
parameters; i.e., in all  regimes \cite{moya}; however, the
solutions are not general because the set of eigenstates that may
be found can not expand all possible (general) states.

It has been shown already that for low intensities it is possible
also to consider the ion micromotion \cite{Schleich}, and by using
Ermakov-Lewis invariant methods \cite{Lewis,Manueli,Manuelii} it
was possible to {\it linearize} the ion-laser Hamiltonian when the
micromotion was included \cite{Jose}. Here, we follow Z\'u\~niga
{\it et al.} \cite{zuniga} and show how it is possible to solve
the ion-laser interaction in different regimes, including high
intensity and medium intensity.

\subsection{Different regimes}
In Section 3.1, we showed that the ion-laser Hamiltonian (\ref{3070}) can be casted in the form given by expression (\ref{3080}) by means of the similarity transformation (\ref{3075}). Therefore, we have {\it linearized} the ion-laser interaction in an
exact way, by means of a unitary transformation; i.e., both Hamiltonians, $\hat{H}_{\textrm{ion}}$ and   $\hat{{\mathcal H}}_{\textrm{ion}}$ are
equivalent. In the following, we will neglect the term $\frac{\nu\eta ^{2}}{4}$ because it only represents a constant shift of all the eigenenergies.
Of course, transformation (\ref{3075}) has to be applied to an initial condition for the internal state of the ion and its vibrational motion wave function. Let us assume that we have the initial state
\begin{equation}
|\psi (0)\rangle=|i\alpha\rangle|e\rangle,
\end{equation}
where $|i\alpha\rangle$ is a coherent state, and for simplicity we
take $\alpha$ a real number (to avoid extra phases later, but the
calculation may be done for complex $\alpha$). Then, we have that
the initial wave function associated with the transformed
Hamiltonian (\ref{3080}) is
\begin{equation}
|\tilde{\psi} (0)\rangle=T|\psi(0)\rangle,
\end{equation}
where the transformation $T$ is given in (\ref{3075}).
If we write the initial wave function in terms of $2\times 2$ matrices, we obtain
\begin{equation}
|\tilde{\psi} (0)\rangle=
\frac{1}{\sqrt{2}}e^{i\hat{n}\frac{\pi}{2}}\left(
{\begin{array}{*{20}c}
 \hat{D}^{\dagger}(i\eta/2) &  \hat{D}(i\eta/2)  \\
   -\hat{D}^{\dagger}(i\eta/2) &  \hat{D}(i\eta/2)
\end{array}} \right)
\left( {\begin{array}{*{20}c}
 |i\alpha\rangle  \\
   0
\end{array}} \right)=\frac{1}{\sqrt{2}}\left( {\begin{array}{*{20}c}
|i(\alpha-\eta/2)\rangle  \\
  - |i(\alpha-\eta/2)\rangle
\end{array}} \right), \label{init}
\end{equation}
Thus, we have changed the complicated Hamiltonian (\ref{3070})
by the linear Hamiltonian (\ref{3080}) via a unitary
transformation. The small prize we have to pay, is that in the
initial wave function the coherent state is displaced and the ion
is initially (in the new frame) in a superposition of ground and
excited states.

\subsection{Medium intensity regime (MIR)}
We already considered this case, when we analyzed in Section 3.3 the blue and red side band by means of the intensity.
In this case, the vibrational frequency is of the
order of twice the field intensity (Rabi frequency). We also
consider the Lamb-Dicke regime; i.e., $\eta \ll  1$. For simplicity,
we will set $\delta=0$ to show the different possibilities we have
now. However, it is not difficult to produce effective Hamiltonians
also in the off-resonance case. In this case, the Hamiltonian
(\ref{3080}) may be casted into (\ref{33070})
which is a Hamiltonian that has been extensively studied
\cite{JCM,JSM}; therefore, we will not add more here, except the
fact that for the medium intensity regime the Hamiltonian
(\ref{3070}) may be exactly expressed as a JCM Hamiltonian via a
unitary transformation and the rotating wave approximation, without extra approximations.

\subsection{Low and high intensity regimes}
In the case of the low intensity regime (LIR) the solution has been known already for several years
\cite{vogeli,vogelii}, however, we will treat it with some detail in Section 5. Here, we will show a different method that is also valid for the high intensity regime (HIR).
Just for the matter of qualitative analysis, let us take $\delta=0$. Consider now  $\Omega\ll\nu $ (LIR) or $\Omega\gg\nu $ (HIR)
in equation (\ref{3080}). As this Hamiltonian for $\delta =0$ is
equivalent to the atom-field interaction, we can borrow knowledge
from such interaction: we know that when the field and atomic
transition frequencies are very different (in our case, it is
translated in the equation $|\nu-2\Omega|\ll \eta\nu/2$, that may
happen in either of both regimes, HIR or LIR), atom and field stop
to exchange energy and we obtain a dispersive Hamiltonian
\cite{Nemes}. The same  happens in the ion-laser interaction, and
via an small rotation approach \cite{klimov}, we will be able to
cast Hamiltonian (\ref{3080}) as an effective (dispersive)
Hamiltonian. \\
By transforming the Hamiltonian (\ref{3080}) with the unitary operators
\begin{equation}
\hat{U}_1=
e^{\xi_1(\hat{a}^{\dagger}\hat{\sigma}_+-\hat{a}\hat{\sigma}_-)},
\qquad \hat{U}_2=
e^{\xi_2(\hat{a}\hat{\sigma}_+-\hat{a}^{\dagger}\hat{\sigma}_-)};
\end{equation}
i.e.,
\begin{equation}
\hat{{\mathcal H}}_{\textrm{eff}}=\hat{U}_2\hat{U}_1 \hat{{\mathcal
H}}_{\textrm{ion}} \hat{U}_1^{\dagger}\hat{U}_2^{\dagger}
\end{equation}
with $\xi_1,\xi_2 \ll 1$, using
\begin{equation}
\xi_1=\frac{\eta\nu}{2(\nu+2\Omega)}\qquad
\xi_2=\frac{\eta\nu}{2(2\Omega-\nu)},
\end{equation}
and remaining up to first order in
the expansion $e^{\xi A}Be^{-\xi A}=B+\xi
[A,B]+\frac{\xi^2}{2!}[A,[A,B]]+ ...\approx B+\xi [A,B]$ \cite{klimov}, we get the effective Hamiltonian
\begin{equation}\label{effion}
\hat{{\mathcal H}}_{\textrm{eff}}= \nu \hat{a}^{\dagger} \hat{a} + \Omega
\hat{\sigma}_z - \chi_{\textrm{ion}} \hat{\sigma}_z (\hat{a}^{\dagger}
\hat{a}+\frac{1}{2})+\frac{\delta}{2}(\sigma_++\sigma_-) +
\frac{\kappa }{2}\hat{\sigma}_{z}\left( \hat{a}^{\dagger
}+\hat{a}\right),
\end{equation}
that for $\delta=0$ is known as the dispersive Hamiltonian.
Note that, just as in the atom-field case, there is no need  to
transform the (already transformed) initial state (\ref{init}) as
an small rotation has been applied. We can see that in fact
$\xi_1 , \xi_2 \ll 1$ either in the LIR (in this case we have also
to consider $\eta \ll 1$) or in the HIR (no constrain for $\eta$),
which justifies completely the approximation for the above
Hamiltonian. For the resonant case, $\delta=0$, it becomes
diagonal and we can solve it in an easy way. \\
In Figure 5, we show a plot for the probability to find the ion in its excited state
\begin{equation}
    P_e(t)=\langle \psi(0)|\hat{T}^{\dagger}\exp(it\hat{{\mathcal
    H}}_{\textrm{eff}})\hat{T}|e\rangle\langle e|\hat{T}^{\dagger}\exp(-it\hat{{\mathcal H}}_{\textrm{eff}})\hat{T}|\psi(0)\rangle
\end{equation}
\begin{figure}[h!]\label{Figure5}
    \centering
    \includegraphics [width=15cm,height=10cm]{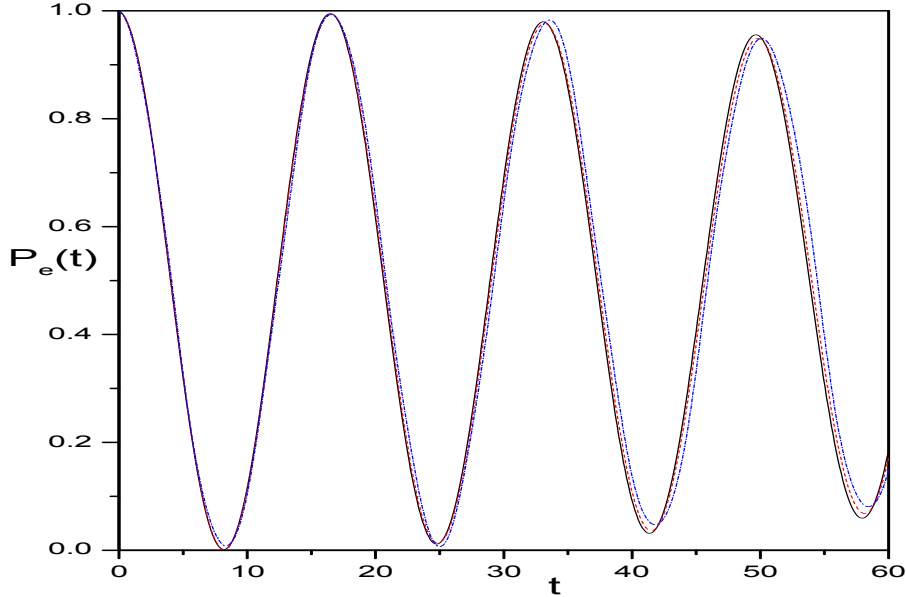}
    \vspace{-1cm}
    \caption {Plot of $P_e(t)$ as a function of $t$ for $k=0$, $\nu=1$, $\Omega=0.2$ and $\eta=0.1$. The vibrational
     motion of the ion is considered to be in a coherent state, $|\alpha|^2=4$ and the ion in its excited state. Solid line
     represents the numerical (exact) solution, dashed line the solution from Section 3 and the dot-dashed line the solution for
     the dispersive Hamiltonian.}
\end{figure}
as a function of time for $k=0$, $\nu=1$, $\Omega=0.2$ and $\eta=0.1$.
The vibrational motion of the ion is considered to be a coherent state $|\alpha|^2=4$,
and the internal state of the ion is excited. The three curves in the figure
correspond to the exact numerical solution (solid line), the
solution from Hamiltonian (\ref{3070})  (dashed line) and the
solution for the dispersive Hamiltonian (\ref{effion}). We can see
excellent agreement among the  three plots for the LIR. Now, for
the HIR in Figure 6, we show a plot also of $P_e(t)$ as a function of
time for the exact numerical solution (solid line) and
our solution from this section (dashed line), but now with the parameters for $k=0$, $\Omega=1$, $\nu=0.2$,
and $\eta=0.1$. Again, it may be noticed the excellent agreement between both curves. We should
stress that there is no other analytical solution to compare with,
as ours is the first analytical solution in this regime (also in the medium intensity regime). \\
\begin{figure}[h!]\label{Figure6}
\centering
\includegraphics[width=15cm,height=10cm]{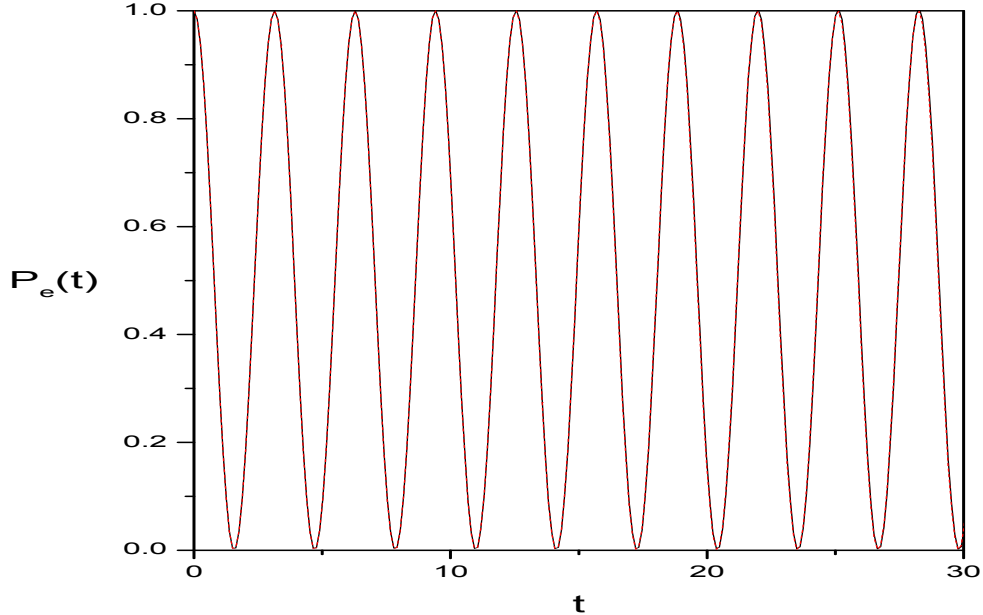}
\vspace{-1cm}
\caption {Plot of $P_e(t)$ as a function of $t$ for $k=0$, $\Omega=1$, $\nu=0.2$ and $\eta=0.1$. The vibrational
motion of the ion is considered to be in a coherent state, $|\alpha|^2=4$ and the ion in its excited state. Solid line
represents the numerical (exact) solution, dashed line the solution for the dispersive Hamiltonian.}
\vspace{1cm}
\end{figure}
The new interaction constants in the effective Hamiltonian (\ref{effion}) have the forms
\begin{equation}
\chi_{\textrm{ion}} =\frac{2\eta ^{2}\nu ^{2}\Omega }{4\Omega ^{2}-\nu
^{2}},\quad \kappa =\frac{\delta \eta \nu ^{2}}{4\Omega ^{2}-\nu^{2}}.
\end{equation}
In the resonant case and high intensity regime, $\Omega \gg \nu$,
it is easy to show that
\begin{equation}
\chi_{\textrm{ion}} \rightarrow \chi^{high} =\frac{2\eta ^{2}\nu
^{2}}{4\Omega }\frac{1}{1-\frac{\nu^{2}}{4\Omega^{2} } } \approx
\frac{\eta^2\nu^2}{2\Omega},
\end{equation}
while in the low intensity regime, $\Omega \ll \nu$, we will have
the same Hamiltonian, but $\chi$ will change to
\begin{equation}
\chi_{\textrm{ion}} \rightarrow \chi_{\textrm{low}} = -2\eta^2 \Omega
\frac{1}{1-\frac{4\Omega^2}{\nu^2}}\approx -2\eta^2\Omega.
\end{equation}
If in Equation (\ref{effion}) we take the detuning $\delta$ different
from zero, we could get the usual blue and red side-bands
interactions (see for instance Ref. \cite{tomb}).
This is done by choosing the value  $\delta=\pm \nu$. The only case in
which we can obtain such regimes is the low intensity case, where
one can perform the rotating wave approximation to the Hamiltonian (\ref{effion}), which
agrees with the usual procedure for obtaining such blue and red
side-band regimes. The high intensity case, $\Omega \gg \nu $, does
not allow such side-bands because in the Hamiltonian
(\ref{effion}) the interaction constants multiplying the different
terms may be of the same order. \\
\\
\\
\section{Low intensity regime}
If we consider that the Hamiltonian (\ref{3020}) corresponds to the
wave function $|\xi(t)\rangle$, the Schr\"{o}dinger equation can
be written as
\begin{equation}\label{80390}
     i \frac{\partial}{\partial t} | \xi \rangle = H | \xi \rangle.
\end{equation}
Let us examine the transformation to a rotating frame of frequency
$\omega$, by means of the unitary transformation
\begin{equation}\label{80380}
     T(t)= \exp{\left( i \frac{\omega}{2} \sigma_z t\right)}.
\end{equation}
Applying the transformation $T$, the wave function
$|\xi(t)\rangle$ transforms in the wave function
$|\phi(t)\rangle$; i.e.,
\begin{equation}\label{80400}
     T(t) | \xi (t) \rangle =  | \phi (t) \rangle,
\end{equation}
and the Hamiltonian transforms in
\begin{equation}\label{80410}
    H_\textrm{T} = i \frac{\partial T (t)}{\partial t} T^{\dagger}(t)  +  T(t) H  T^\dagger (t).
\end{equation}
Writing the position operator $\hat{x}$ in terms of the ladder
operators, expressions (\ref{3030}) and (\ref{3040}), using the
Baker-Hausdorff formula \cite{Schleich,Louisel}, and the commutators of the Pauli
matrices (\ref{pauli}), the explicit transformed Hamiltonian
is
\begin{equation}\label{80420}
    H_\textrm{T} =  \nu \hat{n} + \nu \frac{k}{2} \sigma_z + \lambda E_0 \left[ e^{i\eta (\hat{a}^\dagger + \hat{a})} \sigma_+ + \textrm{H.C.}  \right]
\end{equation}
where $\eta$ is the Lamb-Dicke parameter. The quantity
\begin{equation}\label{80440}
    k \nu = \omega_{21} - \omega
\end{equation}
is the detuning between the plane wave frequency and the transition frequency of the ion; in other words, we are considering that the detuning is a multiple integer of the vibrational frequency of the ion. \\
We need now to factorize the exponential in the Hamiltonian
(\ref{80420}). As $[\hat{a},[\hat{a},\hat{a}^\dagger]]=0$, and
$[\hat{a}^\dagger,[\hat{a},\hat{a}^\dagger]]=0$, we can use the
Baker-Hausdorff formula, and write
\begin{equation}\label{80480}
    e^{-i\eta(\hat{a}+\hat{a}^\dagger)}= e^{-\eta^2/2} e^{- i \eta \hat{a}^\dagger} e^{- i \eta \hat{a}}.
\end{equation}
Expanding in Taylor series the exponentials that contains the
operators, and substituting in the Hamiltonian (\ref{80420}), we
obtain
\begin{equation}\label{80490}
    H_\textrm{T} = \nu \hat{n} + \nu \frac{k}{2} \sigma_z +
     \lambda E_0 e^{-\eta^2/2}\left[ \sigma_- \sum_{n,m=0}^\infty \frac{(-i\eta)^{n+m}}{n!m!}(\hat{a}^\dagger)^n \hat{a}^m + \textrm{H.C.}  \right].
\end{equation}
We go now to the interaction picture, using the transformation
\begin{equation}\label{80500}
    T_{\textrm{free}}=\exp{[it(\nu \hat{n} + \frac{k \nu}{2}\sigma_z)]}.
\end{equation}
We apply this transformation, using the following two commutators,
\begin{equation}\label{80510}
    [\hat{a},f(\hat{a},\hat{a}^\dagger)]=\frac{\partial f}{\partial \hat{a}^\dagger}
\end{equation}
and
\begin{equation}\label{80520}
    [\hat{a}^\dagger,f(\hat{a},\hat{a}^\dagger)]=-\frac{\partial f}{\partial \hat{a}},
\end{equation}
and we obtain
\begin{equation}\label{80530}
    H_\textrm{int} = \Omega e^{-\eta^2/2} \left[ \sigma_- \sum_{n,m=0}^\infty \frac{(-i\eta)^{n+m}}{n!m!}(\hat{a}^\dagger)^n \hat{a}^m e^{-i(n-m-k)\nu t} + \textrm{H.C.}
    \right],
\end{equation}
where $    \Omega=\lambda E_0$ is the exchange energy frequency of the internal and vibrational states,
called Rabi frequency. \\
The interaction Hamiltonian (\ref{80530}) has a diversity of contributions,
each contribution  oscillates with a frequency that is a multiple integer of $\nu$.
We apply now the rotating wave approximation; as the
Schr\"{o}dinger equation is a first order differential equation in
time, we have to integrate it once with respect to time; this
integration brings the sum and the difference of the frequencies
to the denominator. The terms changing slowly will dominate over
the terms changing very fast; so the contribution to the
Hamiltonian of those fast terms is neglected, and only the slowly
changing terms are kept. In this case, the terms that do not
rotate quickly are those whose exponent satisfies the relation
$n-m=k$, and as we already explained, are those terms that we will
keep. This approximation is valid for
\begin{equation}\label{80550}
    \Omega \ll \nu.
\end{equation}
As $\Omega$ is proportional to the amplitude of the laser electric field, from (\ref{80550})
it is clear that this approximation is valid for low intensity. We have then,
\begin{equation}\label{80560}
    H_\textrm{int} = \Omega e^{-\eta^2/2} \left[ \sigma_- \sum_{m=0}^\infty \frac{(-i\eta)^{2m+k}}{(m+k)!m!}(\hat{a}^\dagger)^k (\hat{a}^\dagger)^m \hat{a}^m  + \textrm{H.C.}  \right].
\end{equation}
Using now the fact that the number states is a complete set,
\begin{equation}\label{80570}
    I=\sum_{n=0}^\infty |n\rangle \langle n |,
\end{equation}
where $I$ is the identity operator, we can write
\begin{equation}\label{80580}
    (\hat{a}^\dagger)^m \hat{a}^m= (\hat{a}^\dagger)^m \hat{a}^m \sum_{j=0}^\infty |j\rangle \langle j |= \frac{\hat{n}!}{(\hat{n}-m)!}\sum_{j=0}^\infty |j\rangle \langle j |= \frac{\hat{n}!}{(\hat{n}-m)!},
\end{equation}
that substituted in the Hamiltonian (\ref{80560}), gives us
\begin{equation}\label{80585}
    H_\textrm{int} = \Omega e^{-\eta^2/2} \left[ \sigma_- (\hat{a}^\dagger)^k (-i\eta)^{k}
    \sum_{m=0}^\infty \frac{(-i\eta)^{2m}}{(m+k)!m!} \frac{\hat{n}!}{(\hat{n}-m)!}  + \textrm{H.C.}  \right].
\end{equation}
Using the explicit expression for the associated Laguerre polynomials \cite{Abramowitz,Gradshteyn},
\begin{equation*}
    L_n^{(\alpha)}(x)=\sum_{i=0}^n(-1)^i\frac{(n+\alpha)!}{(n-i)!(\alpha+i)!}\frac{x^i}{i!},
\end{equation*}
we can write
\begin{equation}\label{80590}
    H_\textrm{int} = \Omega e^{-\eta^2/2} \left[ \sigma_- (\hat{a}^\dagger)^k (-i\eta)^{k}
     \frac{\hat{n}!}{(\hat{n}+k)!} L_{\hat{n}}^{(k)}(\eta^2)  + \textrm{H.C.}  \right].
\end{equation}

We will consider now processes where only one phonon is exchanged;
that means that we must take $k=1$ in the Hamiltonian
(\ref{80590}). We will consider also that the oscillation
amplitude of the ion is much smaller than the laser frequency;
that is, $\eta \ll 1$; or in other words, we suppose the
Lamb-Dicke regime. With these two considerations, the Hamiltonian
(\ref{80590}) reduces to (the subscript jc stands for
Jaynes-Cummings)
\begin{equation}\label{80600}
    H_\textrm{jc} = -i \eta \Omega ( \hat{a}^\dagger \sigma_- - \hat{a} \sigma_+ ).
\end{equation}
In agreement with the considerations made above, the Hamiltonian (\ref{80600}) describes emission and absorption of one vibrational excitation, when the atom makes electronic transitions. The first term represents the absorption of a vibrational excitation and the transition of the ion from the excited state to the ground state. The second term represents the inverse process; the ion goes from the ground state to the excited state, annihilating one phonon, and the vibrational state decays in one quanta. \\
All this can be clearly seen, if we apply the Hamiltonian
(\ref{80600}) to the correct states. In the first case, we have to
apply the Hamiltonian to the state $|n\rangle |e\rangle$, which
represents $n$ vibrational quanta and the ion in the excited state
$|e\rangle$; we get,
\begin{equation}\label{80610}
    H_\textrm{jc}|n\rangle |e\rangle \propto |n+1\rangle |g\rangle;
\end{equation}
which is the state with $n+1$ vibrational quanta, and the ion in
the ground state. In the second case, the state is given by
$|n+1\rangle |g\rangle$, and when we apply the Hamiltonian we
obtain,
\begin{equation}\label{80620}
    H_\textrm{jc}|n+1\rangle |g\rangle \propto |n\rangle |e\rangle;
\end{equation}
that is the state with $n$ vibrational quanta, and the ion in the excited state. \\

We can repeat all the above procedure now when the laser frequency
is greater than that of the transition
\begin{equation}\label{80630}
    k \nu = \omega - \omega_{21},
\end{equation}
and obtain the Hamiltonian (clearly now, the subscript ajc stands
for anti-Jaynes-Cummings)
\begin{equation}\label{80640}
    H_\textrm{ajc} = -i \eta \Omega ( \hat{a} \sigma_- - \hat{a}^\dagger \sigma_+ ).
\end{equation}
In the first term, we have the annihilation of one quanta from the vibrational motion and the ion internal transition from the excited state to the ground state. In the second term, we have the creation of one vibrational quanta and the internal excitation of the ion from the ground state to the excited state. In Figure 7, we explain why this Hamiltonian is anti-Jaynes-Cummings type. \\
\begin{figure}[h!]\label{Figure7}
   \centering
  \includegraphics[width=0.8\textwidth]{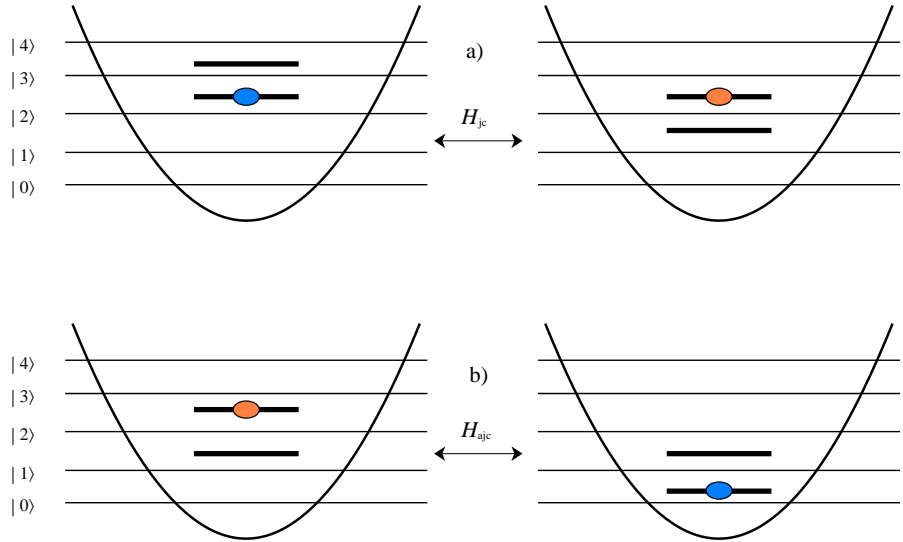}
  \vspace{0.5cm}
  \caption{a) A Jaynes-Cummings Hamiltonian implies the ascent (descent) of one ion vibrational quantum, and at the same time, the transition from an excited (ground) internal state to the ground (excited) state.
  b) An anti-Jaynes-Cummings Hamiltonian annihilate (creates) one quantum from the vibrational motion and transfers the ion internally from the excited (ground) state to the ground (excited) state.}
\end{figure}
Applying the anti-Jaynes-Cummings Hamiltonian to the adequate
states, the previous comments can be easily understood. If we
apply the Hamiltonian (\ref{80640}) to the state $|n\rangle
|e\rangle$, we get
\begin{equation}\label{80650}
    H_\textrm{ajc} |n\rangle |e\rangle= |n-1\rangle |g\rangle,
\end{equation}
and if we apply it to the state $|n-1\rangle |g\rangle$, we get
\begin{equation}\label{80660}
    H_\textrm{ajc} |n-1\rangle |g\rangle= |n\rangle |e\rangle.
\end{equation}

>From the point of view of the trapped ion, all this means that we can take it to its lowest energy vibrational state, alternating successively, and as many times as necessary, the detuning between the frequency of the plane wave and the internal frequency of the ion. Again, we can illustrate all this by applying the correct Hamiltonian to the adequate state. For that let us consider a vibrational state $ |n\rangle$ and the ground internal state; if we apply the Hamiltonian (\ref{80600}), we get
\begin{equation}\label{80670}
     H_\textrm{jc} |n\rangle |g\rangle= |n-1\rangle |e\rangle.
\end{equation}
We apply now the Hamiltonian (\ref{80640}), obtaining
\begin{equation}\label{80680}
     H_\textrm{ajc} |n-1\rangle |e\rangle= |n-2\rangle |g\rangle,
\end{equation}
and the ion has lost two quanta of vibrational energy. Repeating
successively this procedure, we can arrive to the state $|0\rangle
|g\rangle$.

\subsection{Adding vibrational quanta}
We now show how to generate nonclassical vibrational states in the
low intensity regime. From Equation (\ref{80560}) with $k=2$, we
note that if the Lamb-Dicke parameter is much less than one, $\eta
\ll 1$, we can remain to the lowest order in the sum, such that we
obtain the so-called two-phonon Hamiltonian
\begin{equation}\label{twophonon}
      H_{I}  =  \epsilon [\sigma_- {\hat{a}}^2 + \sigma_+{\hat{a}}^{\dagger 2}] ,
\end{equation}
with $\epsilon = -\Omega/2$. For the study of the dynamics of
interest, we need the time evolution described by the Hamiltonian
(\ref{twophonon}).  The advantage of the interactions of
Jaynes-Cummings type consists in the fact that the Hamiltonian can
easily be diagonalized, and using the same procedure that was already used, it is possible to show that
\begin{eqnarray} \label{univib}
\hat{U}_{I}(t) && = \sum_{m=0,1} |m\rangle |g\rangle \langle m| \langle g|\nonumber \\
&&+ \sum_{n=0}^\infty \left[\cos\left (\frac{1}{2} \Omega_n
t\right)
(|n+2\rangle |g\rangle \langle n+2| \langle g | + |n\rangle |e\rangle \langle n| \langle e |)\right.\nonumber\\
&&\left. -i \sin\left (\frac{1}{2} \Omega_n t\right) (|n+2\rangle
|g\rangle \langle n| \langle e |+ |n\rangle |e\rangle \langle
n+2|\langle g |)\right].
\end{eqnarray}
 The quantity $\Omega_n$ is the two-phonon Rabi frequency which is given by
\begin{equation}
\Omega_n = 2 \epsilon \sqrt{(n+1)(n+2)}. \label{rabivib}
\end{equation}
Using these results, the time evolution of the quantum state in
the interaction picture is easily derived for arbitrarily chosen
initial conditions. We have
\begin{equation}
|\Psi(t)\rangle = \hat{U}_{I}(t) |\Psi(0)\rangle. \label{initvib}
\end{equation}
If we consider as initial state the ion in its excited state
$|e\rangle$ and the vibrational state a coherent state, we can
find the atomic inversion, (that we recall that it is defined as the probability to find the ion in
its excited state minus the probability to find it in the ground
state). Using (\ref{univib}) and (\ref{initvib}), we get
\begin{equation}
W(t) = \exp \left( -|\alpha|^2 \right)
\sum_{n=0}^{\infty}\frac{|\alpha|^{2n}}{n!}\cos(2\Omega_n t).
\end{equation}
We plot this function in Figure 8 as a function of the scaled
time $\tau=\epsilon t$. The interaction gives rise to a
quasi-regular evolution of the atomic inversion, unlike the case
of one phonon resonance. This can be used for several purposes,
among them, to add excitations to the vibrational state.
\begin{figure}[h!]\label{Figure8}
  \centering
  \includegraphics[width=0.9\textwidth,height=0.6\textwidth]{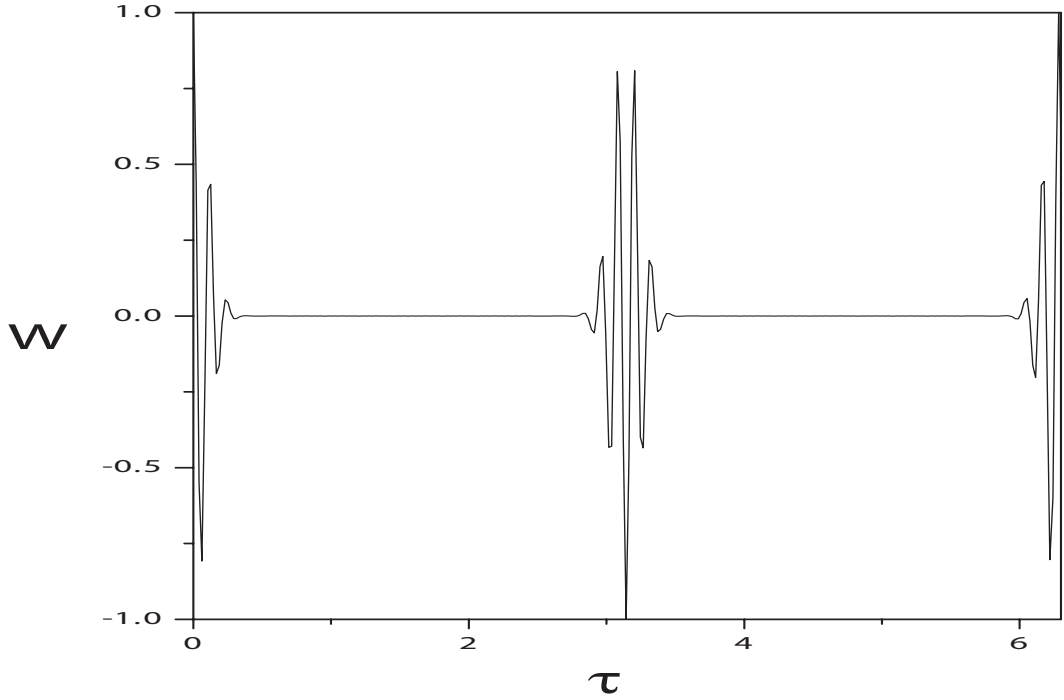}
  \caption{Plot of  the atomic inversion, $W(t)=
  P_2-P_1$, the probability to find the ion in its excited state
  minus the probability to find it in the ground state, for an ion
  initially in its excited state and the vibrational state in a
   coherent state, with $\alpha = 5$. }
\end{figure}
\begin{figure}[h!]\label{Figure9}
   \centering
     \includegraphics [width=0.95\textwidth,height=0.6\textwidth] {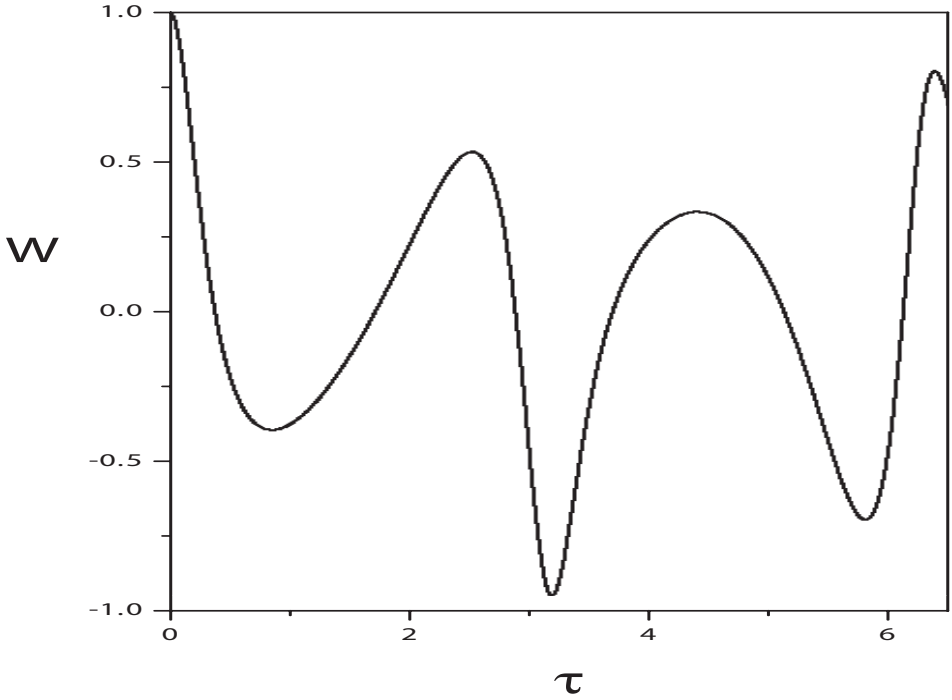}
      \caption{Plot of  the atomic inversion, $W(t)=
       P_2-P_1$, the probability to find the ion in its excited state
       minus the probability to find it in the ground state, for an ion
       initially in its excited state and the vibrational state in a
        thermal distribution with $\bar{n}=2$. }
\end{figure}
In Figure 8, it can be seen that if initially the ion is in its excited
state $|e\rangle$, after an interaction time
$\tau=\pi/\epsilon$ the ion  ends up in its ground state
$|g\rangle$, giving all its energy to the vibrational state by
adding two vibrational quanta. Moreover, the effect of having the
ion in its excited state and after an interaction time having it
in the ground state is shared by all vibrational states, not only
when it is prepared in a coherent state. To illustrate this fact,
we show in Figure 9 the atomic inversion for a thermal distribution
\begin{equation}\label{1d}
   \rho(0)=\sum_{n=0}^{\infty}\frac{\bar{n}^n}{(\bar{n}+1)^{n+1}}  |n\rangle\langle n  |,
\end{equation}
with $\bar{n}$ the average number of thermal phonons. \\
Consider again the initial state
\begin{equation}\label{1d}
    |\Psi(0)\rangle =  |\alpha\rangle  |e\rangle  .
\end{equation}
Combining Equations (\ref{univib}) and (\ref{1d}), and using a
compact operator representation of the Jaynes-Cummings dynamics,
we arrive at
\begin{equation}
|\Psi(t)\rangle = \cos\left(\epsilon t\sqrt{\hat{a}^2
(\hat{a}^\dagger)^2}\right)|\alpha\rangle|e\rangle -
i(\hat{V}^\dagger)^2 \sin\left(\epsilon t\sqrt{\hat{a}^2
(\hat{a}^\dagger)^2}\right)|\alpha\rangle|g\rangle. \label{psivib}
\end{equation}
In order to derive illustrative analytical results, in the
following we will apply the approximation
\begin{equation}\label{aproxtwo}
\sqrt{\hat{a}^2 (\hat{a}^\dagger)^2}\approx \hat{n}+\frac{3}{2};
\end{equation}
although this approximation represents a Taylor-series expansion
for large eigenvalues $n$ of the operator $\hat{n}$, the error is
already small for small $n$-values. For example, for $n=1$ the
relative
error is only $0.02$.  \\
Based on this approximation, one may simplify Equation
(\ref{psivib}) as
\begin{equation}\label{anpsivib}
   |\Psi(t)\rangle \approx  \cos [ \lambda t(\hat{n}+3/2) ]\, |\alpha\rangle|2\rangle
   -i(\hat{V}^\dagger)^2 \sin [ \lambda t(\hat{n}+3/2)]\,  |\alpha\rangle|1\rangle.
\end{equation}
Choosing a particular interaction time $t=\tau$, according to
\begin{equation}
\tau=\pi/\lambda, \label{tau}
\end{equation}
we obtain for the vibrational state vector
\begin{equation}
|\Psi^{(1)}_+(\tau)\rangle_v \approx i(\hat{V}^\dagger)^2
|-\alpha\rangle, \label{psi1int}
\end{equation}
where we have introduced the subscript ''$v$'' (vibrational) to
note that we are not taking into account anymore the  state
$|g\rangle$. Moreover, the subscript ''+'' and the superscript
''$(k)$'' are used to indicate the process of adding (two)
vibrational quanta and the number of such interactions
respectively. \\
The quantum state (\ref{psi1int}) may serve as the initial state
for a second interaction with an ion that is prepared in the same
manner as the first one.  For the same interaction time, $t=\tau$,
after the second interaction (that is completed at time $2\tau$),
the vibrational  state is
\begin{equation}
|\Psi^{(2)}_+(\tau)\rangle_{v} = -(\hat{V}^\dagger)^4
|\alpha\rangle. \label{psi2int}
\end{equation}
By repeating the process $k$ times, one finally obtains for the
quantum state, at the time $t_k$, after completing $k$
interactions,
\begin{equation}\label{psikint}
|\Psi^{(k)}_+(\tau)\rangle_{v} =
i^k(\hat{V}^\dagger)^{2k}|(-1)^k\alpha\rangle.
\end{equation}
After many interactions, the state
$|\Psi^{(k)}_+(\tau)\rangle_{v}$ exhibits a strong sub-Poissonian
character, because while one is adding two excitations per
interaction, at the same time one is
keeping the width of the distribution constant. \\
The excitation distribution $P^{(k)}_n$, after $k$ interactions,
is easily found to be related to the number statistics $P^{(0)}_n$
of the initial state $|\alpha\rangle$ via
\begin{equation}
P^{(k)}_n = P^{(0)}_{n-2k}. \label{distk}
\end{equation}
This result clearly shows that the number statistics is only
shifted but retains its form. \\
One way of studying the properties of the states being generated
is through the Mandel $Q$-parameter \cite{Mandel}, which is defined
by
\begin{equation}
Q=\frac{\left\langle \hat{n}^{2}\right\rangle -\left\langle
\hat{n} \right\rangle ^{2}}{\left\langle \hat{n}\right\rangle }-1,
\label{5.26}
\end{equation}
and where
\begin{equation}
\text{if}\hspace{2mm}Q\hspace{2mm}\left\{
\begin{array}{ll}
>0, & \text{super Poissonian distribution} \\
=0, & \text{Poissonian distribution (coherent state)} \\
<0, & \text{sub-Poissonian} \\
=-1, & \text{number state}.
\end{array}
\right.
\end{equation}
In this case the Mandel $Q$-parameter  is given by
\begin{equation}\label{Mandelp}
Q =  \frac{|\alpha|^2}{|\alpha|^2+2k}-1 ,
\end{equation}
and as the number of interactions increases, the Mandel
$Q$-parameter approaches the value $-1$; i.e., the state acquires
maximum sub-Poissonian character. The sub-Poissonian effect of the
vibrational wave function becomes more significant with increasing
number of interactions.
\subsection{Subtracting vibrational quanta}
It is straightforward to show that, opposite to the case in which we add two phonons per interaction
by initially having the ion in its excited state, two phonons may be removed per interaction by initially
having the ion in its ground state. We therefore expect that instead of squeezing the phonon number
distribution (giving  a sub-Poissonian character to the distribution), in the case in which we subtract phonons, it should be broadened; however, the conjugate variable to the number operator; i.e., the phase operator \cite{Barnett} should have less fluctuations;
a squeezing of the phase distribution then will happen. It is not difficult to show that the phonon distribution when we subtract two phonons per interaction will be given by \cite{moya99}
\begin{equation}
P^{(k)}_n = P^{(0)}_{n+2k}, \label{dist-k}
\end{equation}
from which we can calculate the phase distribution plotted in Figure 10 for several number of interactions.
\begin{figure}[h!]\label{Figure10}
   \centering
     \includegraphics [width=0.9\textwidth] {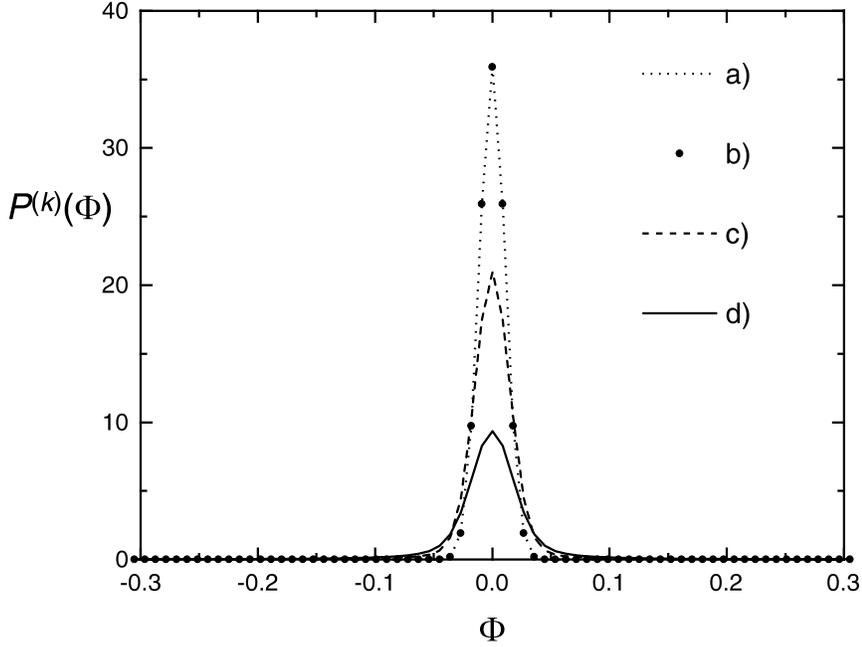}
     \vspace{-1cm}
      \caption{Subtracted phase distribution for an initial coherent
       state $\alpha=45$ and (a) $k = 0$, (b) $k = 900$, (c) $k = 1000$ and (d) $k = 1012.$}
       \end{figure}
\subsection{Filtering specific superpositions of number states}
If instead of one laser, we assume two lasers driving the ion, the
first tuned to the $j$th lower sideband and the second tuned to
the $m$th lower sideband, we may write $ E^{(-)}(\hat{x},t)$ as
\begin{equation}\label{2}
E^{(-)}(\hat{x},t)= E_{j}e^{-i(k_{j}\hat{x}-\omega_{21}+j\nu)t}
+E_{m}e^{-i(k_m\hat{x}-\omega_{21}+m\nu)t},
\end{equation}
where if $m=0$, it would correspond to the driving field being on
resonance with the electronic transition. The position operator
$\hat{x}$ may be written as before,
\begin{equation}\label{3}
k_s\hat{x}=\eta_s(\hat{a}+\hat{a}^\dagger),
\end{equation}
where $k_s, s=j,m$ are the wave vectors of the driving fields and
\begin{equation}
\eta_s=2\pi\frac{\sqrt{\langle
0|\Delta\hat{x}^2|0\rangle}}{\lambda_s}
\end{equation}
are the Lamb-Dicke parameters with $s=j,m$. \\
In the resolved sideband limit, the vibrational frequency $\nu$ is
much larger than other characteristic frequencies, and the
interaction of the ion with the two lasers can be treated
separately using a nonlinear Hamiltonian \cite{vogeli}. The
Hamiltonian (\ref{80590}) in the interaction picture can then be
written as
\begin{equation}\begin{split}\label{4}
      &\hat{H}_\textrm{I} =  \\
      &\sigma_+  \left[  \Omega_j   e^{-\frac{\eta_{j}^2}{2}}
      \frac{\hat{n}! (i \eta)^j}{(\hat{n}+j)!} L_{\hat{n}}^{(j)}(\eta_{j}^2)\hat{a}^j
      +\Omega_m   e^{-\frac{\eta_{m}^2}{2}}
      \frac{\hat{n}!(i \eta)^m}{(\hat{n}+m)!}  L_{\hat{n}}^{(m)}(\eta_{m}^2)\hat{a}^m  \right]  + \textrm{H.C.},
\end{split}\end{equation}
where $L_{\hat{n}}^{(k)}(\eta_{k}^2)$ are the operator-valued
associated Laguerre polynomials, the $\Omega$'s are the Rabi
frequencies and $\hat{n}= \hat{a}^\dagger\hat{a}$. The master
equation which describes this system can be written as
\begin{equation}\label{5}
      \frac{ \partial \hat{\rho}}{\partial t} = - i [\hat{H}_\textrm{I},\hat{\rho}] +
      \frac{\Gamma}{2} \left( 2\sigma_+ \hat{\tilde{\rho}} \sigma_-
         - \sigma_z \hat{\rho} - \hat{\rho} \sigma_z \right)
\end{equation}
where the last term describes spontaneous emission with energy
relaxation rate $\Gamma$, and
\begin{equation}\label{6}
 \hat{\tilde{\rho}}=\frac{1}{2}\int^1_{-1}ds W(s)e^{is\eta_E\hat{x}}\hat{\rho} e^{-is\eta_E\hat{x}}
\end{equation}
accounts for changes of the vibrational energy because of
spontaneous emission. Here $\eta_E$ is the Lamb-Dicke parameter
corresponding to the field (\ref{2}) and $W(s)$ is the angular
distribution of
spontaneous emission \cite{vogeli}. \\
The steady-state solution to Equation (\ref{5}) is obtained by
setting $\partial \hat{\rho}/\partial t=0$, and may be written as
\begin{equation}\label{7}
 \hat{\rho}_s= | \psi_s\rangle |g\rangle    \langle\psi_s|\langle g|,
\end{equation}
where $|g\rangle$ is the electronic ground state and
$|\psi_s\rangle$ is the vibrational steady-state of the ion, given
by
\begin{equation}\begin{split}\label{8}
    \left[\Omega_j e^{-\eta_{j}^2/2} \frac{\hat{n}!(i \eta_j)^j}{(\hat{n}+j)!}   L_{\hat{n}}^{(j)}(\eta_{j}^2)\hat{a}^j
    +\Omega_m e^{-\eta_{k}^2/2} \frac{\hat{n}!(i \eta_m)^m}{(\hat{n}+m)!}  L_{\hat{n}}^{(m)}(\eta_{m}^2)\hat{a}^m  \right] |\psi_s\rangle=0.
\end{split}\end{equation}
For simplicity, we will concentrate in the $j=1$ and $m=0$ case
(single number state spacing) for which Equation (\ref{8}) is
written as
\begin{equation}\label{9}
    \left[i\Omega_1 \eta_1 e^{-\eta_{1}^2/2}
    \frac{L_{\hat{n}}^{(1)}(\eta_{1}^2)}{\hat{n}+1}\hat{a} +\Omega_0
     e^{-\eta_{0}^2/2} L_{\hat{n}}(\eta_{0}^2)\right] |\psi_s\rangle=0.
\end{equation}
Note that $\hat{H}_\textrm{I}|1\rangle|\psi_s\rangle=0$, so that
ion and laser have stopped to interact, which occurs when the ion
stops to fluoresce. For the $j=1$ and $k=0$ case, and assuming
$L_k^{(1)}(\eta_1^2)\neq 0$ and $L_k(\eta_0^2)\neq 0$ for all $k$,
one generates nonlinear coherent states \cite{vogeliv}. However, by
setting a value to one of the Lamb-Dicke parameters such that, for
instance,
\begin{equation} \label{10}
      L_q(\eta_0^2)=0,
\end{equation}
for some integer $q$, we obtain that, by writing $|\psi_s\rangle$
in the number state representation,
\begin{equation}\label{11}
      |\psi_s(\eta_0)\rangle=\frac{1}{N_0}\sum_{n=0}^qC^{(0)}_n|n\rangle,
\end{equation}
(the argument of $\psi_s$ denotes the condition we apply; i.e., in
Equation (\ref{11}), the condition is on $\eta_0$) where
\begin{eqnarray}
  \nonumber C^{(0)}_n &=& \left(-\frac{\Omega_0 e^{-\eta_0^2/2}}{\Omega_1
e^{-\eta_1^2/2}}\right)^n
(n!)^{1/2}\prod_{m=0}^{n-1}\frac{L_m(\eta_0^2)}{L_m^{1}(\eta_1^2)}, \\
   C^{(0)}_0 &=& 1,
\end{eqnarray}
and
\begin{equation}
N_0^2=\sum_{n=0}^q|C_n^{(0)}|^2
\end{equation}
is the normalization constant. \\
If instead of condition (\ref{10}), we choose
\begin{equation}\label{12}
L_p^{(1)}(\eta_1^2)=0,
\end{equation}
we obtain the wave function
\begin{equation}\label{13}
|\psi_s(\eta_1)\rangle=\frac{1}{N_1}\sum_{n=p+1}^\infty
C_n^{(1)}|n\rangle,
\end{equation}
where now
\begin{eqnarray}
  \nonumber C^{(1)}_n &=& \left(-\frac{\Omega_0 e^{-\eta_0^2/2}}{\Omega_1
e^{-\eta_1^2/2}}\right)^{n-p-1}
\sqrt{\frac{n!}{(p+1)!}}\prod_{m=p+1}^{n-1}\frac{L_m(\eta_0^2)}{L_m^{1}(\eta_1^2)}, \\
  C^{(1)}_{p+1} &=& 1,
\end{eqnarray}
and
\begin{equation}
N_1^2=\sum_{n=p+1}^\infty|C_n^{(1)}|^2.
\end{equation}
Combining both conditions, (\ref{10}) and (\ref{12}), one would
obtain for $q>p$,
\begin{equation}
|\psi_s(\eta_0,\eta_1)\rangle=\frac{1}{N_{01}}\sum_{n=p+1}^q
C^{(1)}_n|n\rangle, \label{14}
\end{equation}
with
\begin{equation}
N_{01}^2=\sum_{n=p+1}^q|C_n^{(1)}|^2.
\end{equation}
In this way, by setting the conditions (\ref{10}), (\ref{12}) or
both, we can engineer states in the following three zones of the
Hilbert space: (a)  from $|0\rangle$ to $|q\rangle$, (b) from
$|p+1\rangle$ to $|\infty\rangle$, or (c) from $|p+1\rangle$ to
$|q\rangle$. In the later case, by setting $q=p+1$, generation of
the number state $|q\rangle$ is achieved. \\
We should remark that by selecting further apart sidebands one
would obtain a different spacing in Equations (\ref{11}),
(\ref{13}) and (\ref{14}). For instance, by choosing $j=2$ and
$k=0$ one would obtain only even or odd number states in those
equations (depending in this case on initial conditions, and
$W(s)$, the angular distribution of spontaneous emission). Also,
it should be noticed that one can use the parameters $j=m+1$ and
$k=m$ (with $m \neq 0$) (in the single number state spacing case)
to extend the possibilities of choosing Lamb-Dicke parameters.
Lamb-Dicke parameters of the order of one (or less) are needed
(for conditions (\ref{10}) and (\ref{12})), which can be achieved
by varying the geometry of the lasers. For example, by setting
$\eta_0=1$, we have $L_1(\eta^2_0=1)=0$, and therefore we obtain
the qubit
\begin{equation}\label{15}
|\psi_s(\eta_0=1)\rangle=\frac{1}
{\sqrt{1+|\frac{\Omega_0}{\Omega_1}|^2 e^{\eta_1^2-1}}}
\left(|0\rangle- \frac{\Omega_0 e^{-1/2}}{\Omega_1
e^{-\eta_1^2/2}}|1\rangle\right),
\end{equation}
where by changing the Rabi frequencies, one has control of the
amplitudes. Finally, note that we could have also chosen to drive
the $q$th upper sideband instead of the $k$th lower sideband in
Equation (\ref{2})  with basically the same results.
\begin{figure}[h!]\label{Figure11}
    \centering
      \centering
      \includegraphics[width=0.7\textwidth]{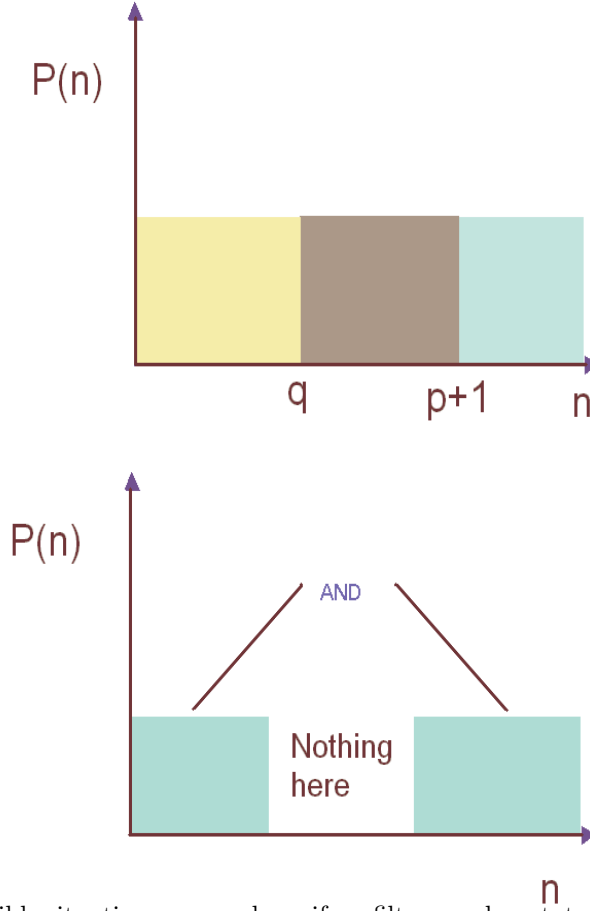}
      \vspace{-2cm}
      \caption{Possible situations we can have if we filter number states with the proposals of this section. }
\end{figure}

\subsection{$N00N$ states}
Nonclassical states have attracted a
great deal of attention in recent years, among them (a)
macroscopic quantum superpositions of quasiclassical coherent
states with different mean phases or amplitudes \cite{1,2}, (b)
squeezed states \cite{loudon,Manueli},  (c) the particularly important limit of
extreme squeezing; i.e., Fock or number states, and more recently,
(d) nonclassical states of combined photon pairs also called $N00N$
states \cite{5,6}. It is well known that these multiphoton
entangled states can be used to obtain high-precision phase
measurements, becoming more and more advantageous as the number of
photons grows. Many applications in quantum imaging, quantum
information and quantum metrology \cite{7} depend on the
availability of entangled photon pairs because entanglement is a
distinctive feature of quantum mechanics that lies at the core of
many new applications. These maximally path-entangled multiphoton
states may be written in the form
\begin{equation}
    \left| {N00N} \right\rangle _{a,b}  = \frac{1}{{\sqrt 2 }}\left(
    {\left| N \right\rangle _a \left| 0 \right\rangle _b  + \left| 0
    \right\rangle _a \left| N \right\rangle _b } \right).
\end{equation}
It has been pointed out that $N00N$ states manifest unique coherence
properties by showing that they exhibit a periodic transition
between spatially bunched and antibunched states when undergo
Bloch oscillations. The period of the bunching/antibunching
oscillation is $N$ times faster than the period of the oscillation
of the photon density \cite{12}. \\
The greatest $N$ for which $N00N$ states have been produced is
$N=5$ \cite{5}. Most  schemes to generate  this class of states are
either for optical \cite{5,6} or microwave \cite{Diana} fields. In
this contribution, we would like to analyze the possibility to
generate them in ions \cite{wine3,ion,Buzek,tomb,moya}; i.e., $N00N$
states of their vibrational motion. We will show that they may be
generated with $N=8$.

\subsubsection{Ion vibrating in two dimensions}
We consider an ion in a two-dimensional Paul trap \cite{Zheng}, and we assume that the ion is driven by a plane wave
\begin{equation}
    E^{(-)}(\hat{x},\hat{y},t)=E_0e^{-i(k_x \hat{x}+k_y \hat{y}+\omega)t},
\end{equation}
with $k_j, j=x,y$ the wavevectors of the driving field. The Hamiltonian has the form
\begin{equation}\begin{split}\label{ion0001}
    H =& \nu _x \hat a_x^ {\dagger}  \hat a_x  + \nu _y \hat a_y^ {\dagger}  \hat a_y  + \frac{{\omega _{21} }}{2}\hat \sigma _z \\
    +& \Omega _x \left\{ {e^{ - i\left[ {\eta _x \left( {\hat a_x  + \hat a_x^ {\dagger}  } \right) + \omega t} \right]} \hat \sigma _ +   + \textrm{H.C.}} \right\}
    + \Omega _y \left\{ {e^{ - i\left[ {\eta _y \left( {\hat a_y  + \hat a_y^ {\dagger}  } \right) + \omega t} \right]} \hat \sigma _ +   +\textrm{H.C.}} \right\}.
\end{split}\end{equation}
where we have defined the Lamb-Dicke parameters
\begin{equation}
\eta_x=2\pi\frac{\sqrt{_x\langle 0|\Delta\hat{x}^2|0\rangle}_x}{\lambda_x}, \qquad
\eta_y=2\pi\frac{\sqrt{_y\langle 0|\Delta\hat{y}^2|0\rangle_y}}{\lambda_y},
\end{equation}
and redefined the ladder operators according to
\begin{equation}
k_x \hat{x}=\eta_x(a_x+a_x^{\dagger}), \qquad k_  \textrm{ and }  \hat{y}=\eta_y(a_y+a_y^{\dagger}).
\end{equation}
In the resolved sideband limit, the vibrational frequencies
$\nu_x$ and $\nu_y$ are much larger than other characteristic
frequencies and the interaction of the ion with the two lasers can
be treated separately using a nonlinear Hamiltonian \cite{vogeli,vogelii}. \\
We consider that the ion is trapped in the $x$ - axis; i.e., $\Omega_x\neq0$ and $\Omega_y=0$; then
\begin{equation}
H_x=\nu _x \hat a_x^ {\dagger}  \hat a_x  + \frac{{\omega _{21} }}{2}\hat \sigma _z  + \Omega _x \left\{ {e^{ - i\left[ {\eta _x \left( {\hat a_x  + \hat a_x^ {\dagger}  } \right) + \omega t} \right]} \hat \sigma _ +  + \textrm{H.C.}}\right\}.
\end{equation}
We write $\omega_{21}=\omega+\delta$, where $\delta$ is the detuning, to obtain
\begin{equation}
H_x = \nu _x \hat a_x^ {\dagger}  \hat a_x  + \frac{{\left( {\omega  + \delta } \right)}}{2}\hat \sigma _z  + \Omega _x \left\{ {e^{ - i\left[ {\eta _x \left( {\hat a_x  + \hat a_x^ {\dagger}  } \right) + \omega t} \right]} \hat \sigma _ +   + \textrm{H.C.}} \right\}.
\end{equation}
We transform the Hamiltonian to a frame rotating at $\omega$ frequency by means of the transformation
\begin{equation}
T=e^{-i\frac{\omega t}{2}\sigma_z},
\end{equation}
and we get
\begin{equation}
H_x  = \nu _x \hat a_x^ {\dagger}  \hat a_x  + \frac{\delta }{2}\hat \sigma _z  + \Omega _x \left\{ {e^{ - i\left[ {\eta _x \left( {\hat a_x  + \hat a_x^ {\dagger}  } \right)} \right]} \hat \sigma _ +   + e^{i\left[ {\eta _x \left( {\hat a_x  + \hat a_x^ {\dagger}  } \right)} \right]} \hat \sigma _ -  } \right\}.
\end{equation}
Using the Baker-Hausdorff formula \cite{Louisel}, and expanding the exponentials in Taylor series, we cast the Hamiltonian to
\begin{equation}
H_x  = \nu _x \hat a_x^ {\dagger}  \hat a_x  + \frac{\delta }{2}\hat \sigma _z  + \Omega _x \left[ {e^{ - \frac{{\eta _x^2 }}{2}} \sum\limits_{n,m} {\frac{{\left( {- i\eta _x } \right)^n }}{{n!}}\frac{{\left( { - i\eta _x } \right)^m }}{{m!}}} \hat a_x^{ \dagger n}\hat a_x^m \hat \sigma _ +   + \textrm{H.C.}} \right].
\end{equation}
Going now to the interaction picture,
\begin{equation}
H_{Ix}  = \Omega _x \left[ {e^{ - \frac{{\eta _x^2 }}{2}} \sum\limits_{n,m} {\frac{{\left( { - i\eta _x } \right)^n }}{{n!}}\frac{{\left( { - i\eta _x } \right)^m }}{{m!}}} \hat a_x^{ \dagger n} \hat a_x^m \hat \sigma _ +  e^{i\nu _x t\left( {n - m + k} \right)}  + \textrm{H.C.}} \right].
\end{equation}
We consider now the low-intensity regime; i.e., $\Omega_x<<\nu_x$, and we apply the rotating wave approximation, to get
\begin{equation}
H_{Ix}  = \Omega _x \left[ {e^{ - \frac{{\eta _x^2 }}{2}} \left( { - i\eta _x } \right)^k \sum\limits_{n = 0}^\infty  {\frac{{\left( { - \eta _x } \right)^{2n} }}{{n!\left( {k + n} \right)!}}} \hat a_x^{ \dagger n} \hat a_x^{k + n} \hat \sigma _ +   + \textrm{H.C.}} \right],
\end{equation}
by substituting $\hat a_x^{ \dagger n} \hat a_x^n  = \frac{{\hat n!}}{{\left( {\hat n - n} \right)!}}$, multiplying by
$\frac{{\left( {\hat n + k} \right)!}}{{\left( {\hat n + k} \right)!}}$, and rearranging terms
\begin{equation}
H_{Ix}  = \Omega _x \left[ {e^{ - \frac{{\eta _x^2 }}{2}} \left( { - i\eta _x } \right)^k \frac{{\hat n!}}{{\left( {\hat n +k} \right)!}}L_{\hat n}^k \left( {\eta _x^2 } \right)\hat a_x^k \hat \sigma _ +   + \textrm{H.C.}} \right],
\end{equation}
 where we have identified
 $L_{\hat n}^k \left( {\eta _x^2 } \right) = \sum\limits_{n = 0}^{\hat n} {\frac{{\left( { - 1} \right)^n \left( {\eta _x^2 } \right)^n }}{{n!}}\frac{{\left( {\hat n + k} \right)!}}{{\left( {n + k} \right)!\left( {\hat n - n} \right)!}}}$,
 with the associated Laguerre polynomials;  so that finally
\begin{equation}\label{71400}
H_{Ix}  = \Omega _x \left( f_x^k \left( {\hat n} \right)\hat a_x^k \hat \sigma _ +   + \hat a_x^{ \dagger k} f_x^{*k} \left( {\hat n} \right)\hat \sigma _ - \right),
\end{equation}
where
\begin{equation}
     f_x^k \left( {\hat n} \right) = e^{ - \frac{{\eta _x^2 }}{2}} \left( { - i\eta _x } \right)^k \frac{{\hat n!}}{{\left( {\hat n + k} \right)!}}L_{\hat n}^k \left( {\eta _x^2 } \right).
\end{equation}
We write the Hamiltonian, given by expression (\ref{71400}), in the following matrix form
\begin{equation}
H_{Ix}  = \left( {\begin{array}{*{20}c}
   0 & {\Omega _x f_x^k \left( {\hat n} \right)\hat a_x^k }  \\
   {\Omega _x \hat a_x^{ \dagger k} f_x^{*k} \left( {\hat n} \right)} & 0  \\
\end{array}} \right),
\end{equation}
which, by using the Susskind-Glogower phase operator \cite{susskind}, we can be written as
\begin{equation}
H_{Ix}  = \left( {\begin{array}{*{20}c}
   1 & 0  \\
   0 & {\hat V_x^{ \dagger k} }  \\
\end{array}} \right)
H_{1x}
\left( {\begin{array}{*{20}c}
   1 & 0  \\
   0 & {\hat V_x^k }  \\
\end{array}} \right)
\end{equation}
where we have introduced and defined
\begin{equation}
H_{1x}=\left( {\begin{array}{*{20}c}
   0 & {\Omega _x f_x^k \left( {\hat n} \right)\sqrt {\hat a_x^k \hat a_x^{ \dagger k} } }  \\
   {\Omega _x f_x^{*k} \left( {\hat n} \right)\sqrt {\hat a_x^k \hat a_x^{ \dagger k} } } & 0  \\
\end{array}} \right).
\end{equation}
The evolution operator for this last transformed Hamiltonian,  $U_{1x}=e^{-iH_{1x}t}$ , may be calculate easily. For this we need
\begin{equation}
    H_{1x}^{2m}  = \left( {\begin{array}{*{20}c}
   {\Omega _x^{2m} \left| {f_x^k \left( {\hat n} \right)} \right|^{2m} \left( {\sqrt {\hat a_x^k \hat a_x^{ \dagger k} } } \right)^{2m} } & 0  \\
   0 & {\Omega _x^{2m} \left| {f_x^k \left( {\hat n} \right)} \right|^{2m} \left( {\sqrt {\hat a_x^k \hat a_x^{ \dagger k} } } \right)^{2m} }  \\
\end{array}} \right),
\end{equation}
and
\begin{equation}
    H_{1x}^{2m + 1}  = \left( {\begin{array}{*{20}c}
   0 & {\Omega _x^{2m + 1} \frac{{\left| {f_x^k \left( {\hat n} \right)} \right|^{2m + 1} f_x^{*k} \left( {\hat n} \right)}}{{\left| {f_x^k \left( {\hat n} \right)} \right|}}\left( {\sqrt {\hat a_x^k \hat a_x^{ \dagger k} } } \right)^{2m + 1} }  \\
   {\Omega _x^{2m + 1} \frac{{\left| {f_x^k \left( {\hat n} \right)} \right|^{2m + 1} f_x^{*k} \left( {\hat n} \right)}}{{\left| {f_x^k \left( {\hat n} \right)} \right|}}\left( {\sqrt {\hat a_x^k \hat a_x^{ \dagger k} } } \right)^{2m + 1} } & 0 \\
\end{array}} \right).
\end{equation}
Then
\begin{equation}
    U_{1x} \left( t \right) = \sum\limits_m {\frac{{\left( { - it} \right)^{2m} }}{{\left( {2m} \right)!}}} H_{1x}^{2m}  + \sum\limits_m {\frac{{\left( { - it} \right)^{2m + 1} }}{{\left( {2m + 1} \right)!}}} H_{1x}^{2m + 1},
\end{equation}
and therefore
\begin{equation}
U_{Ix} \left( t \right) = \left( {\begin{array}{*{20}c}
   {\cos\left( {\Omega _x t\left| {f_x^k \left( {\hat n} \right)} \right|\sqrt {\hat a_x^k \hat a_x^{ \dagger k} } } \right)} & { - i\left( { - i} \right)^k \sin\left( {\Omega _x t\left| {f_x^k \left( {\hat n} \right)} \right|\sqrt {\hat a_x^k \hat a_x^{ \dagger k} } } \right)\hat V_x^k }  \\
   { - i\hat V_x^{ \dagger k} \left( i \right)^k \sin\left( {\Omega _x t\left| {f_x^k \left( {\hat n} \right)} \right|\sqrt {\hat a_x^k \hat a_x^{ \dagger k} } } \right)} & {\hat V_x^{ \dagger k} \cos\left( {\Omega _x t\left| {f_x^k \left( {\hat n} \right)} \right|\sqrt {\hat a_x^k \hat a_x^{ \dagger k} } } \right)\hat V_x^k }  \\
\end{array}} \right).
\end{equation}
Finally, as $\sqrt{\hat{a}_x^k\hat{a}_x^{\dagger }{}^k}=\sqrt{\frac{(\hat{n}+k)!}{\hat{n}!}}$, we can write
\begin{equation}
    U_{I x}(t)=\left(
    \begin{array}{cc}
       \cos \left[\Omega _xt\left|f_x^k\left(\hat{n}\right)\right|\sqrt{\frac{(\hat{n}+k)!}{\hat{n}!}}\right] & (-i)^{k+1} \sin \left[\Omega _xt\left|f_x^k\left(\hat{n}\right)\right|\sqrt{\frac{(\hat{n}+k)!}{\hat{n}!}}\right]\hat{V}_x^k \\
      -(i)^{k+1} \hat{V}_x^{\dagger ^k}\sin \left[\Omega _xt\left|f_x^k\left(\hat{n}\right)\right|\sqrt{\frac{(\hat{n}+k)!}{\hat{n}!}}\right] & \hat{V}_x^{\dagger ^k}\cos \left[\Omega _xt\left|f_x^k\left(\hat{n}\right)\right|\sqrt{\frac{(\hat{n}+k)!}{\hat{n}!}}\right]\hat{V}_x^k
    \end{array}
\right).
\end{equation}
Now, we consider as initial state of the ion a number state $|n\rangle$ for the vibrational motion and the excited state $|e\rangle$ for the internal states; i.e.,
\begin{equation}
\left| {\psi \left( 0 \right)} \right\rangle  = \left(
{\begin{array}{*{20}c}
   {\left| n  \right\rangle }  \\
   0  \\
\end{array}} \right).
\end{equation}
The probability, after the time $t$, of finding the ion in its internal excited state is then
\begin{equation}
P_e \left( t \right) = \sum _{m=0}^{\infty } \langle e|\langle m|\psi (t)\rangle \langle \psi (t)|m\rangle |e\rangle =  \cos^2 \left( {\Omega _x
t\left| {f_x^k \left( { n} \right)} \right| \sqrt{\frac{(\hat{n}+k)!}{\hat{n}!}} } \right).
\end{equation}
It is clear that after a time
\begin{equation}
    t_0=\frac{\pi }{2}\frac{1}{\Omega _x\left|f_x^k(n)\right|}\sqrt{\frac{n!}{(n+k)!}}
\end{equation}
the probability to find the ion in its internal excited state is 0, so at that time the ion is in its internal ground state with probability 1. This situation is obviously repeated periodically; every $2j+1, j=0,1,2,...$ times $t_0$, the ion will be in its ground state.
We plot in Figure 12 the probability for the case when $k=4$ and the ionic vibration is initially in a number state with $n=4$.
\begin{figure}[h!]\label{Figure12}
        \centering
        \includegraphics [width=0.8\textwidth]{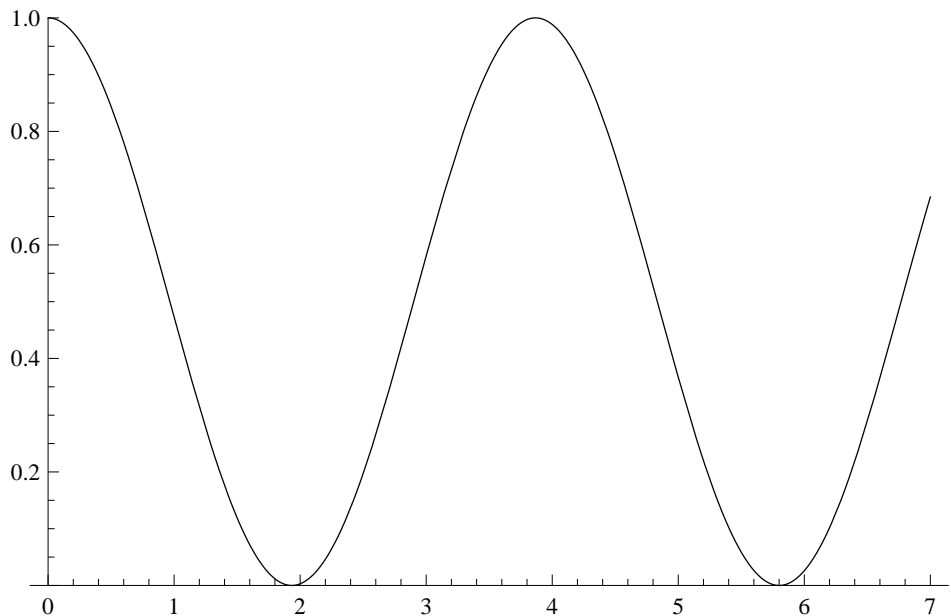}
         \caption {The probability, after the time $\tau$, of finding the ion in its internal excited state when initially the ion was in its excited state and the  ionic vibration was a number state with $n=4$ and $\eta_x=0.1$.}
\end{figure}
As can bee seen from Hamiltonian (\ref{71400}), when the probability of finding the ion in the excited state goes to zero, the ion is giving four phonons to the vibrational motion. Now, if we consider the ion initially in its ground state, the probability to find it in the ground state at the same time $t_0$ is also zero. In this case the ion removes four phonons of the vibrational motion. \\
If we consider now that the ion is trapped in the $y$ - axis; i.e., $\Omega_y\neq0$ and $\Omega_x=0$, we get exactly the same expressions and the same results with the variable $y$ instead of $x$.

\subsubsection{Generation of $N00N$ states}
By  starting with the ion in the excited state and the vibrational state in the vacuum state; i.e.,
$|0\rangle_x|0\rangle_y$, if we set $\eta_y=0$, after the time $\tau_p$ when the probability to find the ion in
its excited state is zero (meaning that the ion, by passing from its excited to its ground state, gives $4$ phonons to the
vibrational motion),  we can generate the state
$|4\rangle_x|0\rangle_y$. Repeating this procedure (with the ion
reset again to the excited state, via a rotation), but now with
$\eta_x=0$, four phonons are added to the $y$-vibrational motion,
generating the two-dimensional state $|4\rangle_x|4\rangle_y$. \\
Therefore, if we consider the ion initially in a superposition of ground and
excited states, and the $|4\rangle_x|4\rangle_y$ vibrational state; i.e.,
\begin{equation}
|\psi_{init}\rangle=\frac{1}{\sqrt{2}}(|e\rangle+|g\rangle)|4\rangle_x|4\rangle_y,
\end{equation}
for $\eta_y=0$ and $t_0$, the state generated is
\begin{equation}
|\psi_{\eta_y=0}\rangle=\frac{i}{\sqrt{2}}(|e\rangle|0\rangle_x+|g\rangle|8\rangle_x)|4\rangle_y.
\end{equation}
Now, we consider this state as initial state for the next interaction with $\eta_x=0$ and still the interaction time $\tau_p$, to produce
\begin{equation}
|\psi_{\eta_x=0}\rangle=-\frac{1}{\sqrt{2}}(|e\rangle|0\rangle_x|8\rangle_y+|g\rangle|8\rangle_x|0\rangle_y).
\end{equation}
Next, the ion is rotated via a classical field (an on-resonance interaction) such that the state
\begin{equation}
|\psi_R\rangle=-\frac{1}{{2}}\left[|e\rangle(|0\rangle_x|8\rangle_y-|8\rangle_x|0\rangle_y)
+|g\rangle(|0\rangle_x|8\rangle_y+|8\rangle_x|0\rangle_y\right]
\end{equation}
is obtained. Finally by measuring the ion in its excited state we produce the $N00N$ state
\begin{equation}
|N00N_e\rangle=\frac{1}{\sqrt{2}}(|0\rangle_x|8\rangle_y-|8\rangle_x|0\rangle_y),
\end{equation}
and if the ion is measure in the ground state, also a $N00N$ state is produced:
\begin{equation}
|N00N_g\rangle=\frac{1}{\sqrt{2}}(|0\rangle_x|8\rangle_y+|8\rangle_x|0\rangle_y),
\end{equation}

\subsection{Measuring squeezing}
In this section, we have shown how to generate nonclassical states of the vibrational motion of the ion. The question arises:
Can we measure nonclassical features in this interaction? Here we propose a method to measure squeezing. In order to achieve this we  need to be able to measure quantities like
\begin{equation}
\langle \hat{X} \rangle = \langle \hat{a} \rangle+ \textrm{c.c.}, \qquad
\langle \hat{X}^2 \rangle = \langle \hat{a}^2 \rangle + \langle
[\hat{a}^{\dagger}]^2\rangle+2\langle \hat{n} \rangle + 1.
\end{equation}
Below we will show how it is possible to measure such quantities by utilizing an atom as a measuring device. Consider the Hamiltonian (109), which in matrix form may be written as
\begin{equation}
{H}=-i\eta\Omega(\hat{a}^{\dagger}\hat{\sigma}_--\hat{\sigma}_+\hat{a})=
-i\eta\Omega\left(
\begin{array}{cc}
0 & -\hat{a}
\\  \hat{a}^{\dagger}& 0
\end{array}
\right). \label{hamiltonian}
\end{equation}

We can re-write Hamiltonian (\ref{hamiltonian}) with the help of Susskind-Glogower operators \cite{susskind} as
\begin{equation}
{H}=-i\eta\Omega{R} \left(
\begin{array}{cc}
0 & -\sqrt{\hat{n}+1}
\\  \sqrt{\hat{n}+1}& 0
\end{array}
\right) {R}^{\dagger},
\end{equation}
where
\begin{equation}
{R} =  \left(
\begin{array}{cc}
1 & 0
\\ 0 & \hat{V}^{\dagger}
\end{array}
\right)
\end{equation}
with $\hat{V}=\frac{1}{\sqrt{\hat{n}+1}}\hat{a}$. Note that ${R}^{\dagger} {R}=1$, but ${R}{R}^{\dagger} \ne 1$.
This allows us to write the evolution operator as
\begin{equation}\label{evol}
{U}(t)= {R} \left(
\begin{array}{cc}
\cos(\eta\Omega t\sqrt{\hat{n}+1}) & -\sin(\eta\Omega
t\sqrt{\hat{n}+1})
\\ \sin(\eta\Omega t\sqrt{\hat{n}+1}) & \cos(\eta\Omega t\sqrt{\hat{n}+1})
\end{array}
\right) {R}^{\dagger}
\end{equation}
We are neglecting a term $|0\rangle\langle 0|$ in the above
evolution operator (in the element "$2,2$"), that however will not
affect the measurement of squeezing as we will consider the ion
in the excited state.  We consider the  vibrational wave function in an unknown state,
such that the initial state of the system is $|\psi(0)\rangle =
|e\rangle|\psi_v(0)\rangle$; the average of the operator $\hat{\sigma}_+$ is given by
\begin{eqnarray}
\langle  \hat{\sigma}_+ \rangle &=& \langle
\psi_v(0)|\cos(\eta\Omega
t\sqrt{\hat{n}+1})\hat{V}^{\dagger}\sin(\eta\Omega
t\sqrt{\hat{n}+1})|\psi_v(0)\rangle
\\ \nonumber &=& \frac{1}{2}\langle \psi_v(0)|\hat{V}^{\dagger}\left(\sin[\eta\Omega
t\hat{\Delta}_+(\hat{n})] - \sin[\eta\Omega
t\hat{\Delta}_-(\hat{n})]\right)|\psi_v(0)\rangle ,\label{sigma}
\end{eqnarray}
where \begin{equation} \hat{\Delta}_+(\hat{n})=
\sqrt{\hat{n}+2}+\sqrt{\hat{n}+1}, \qquad \hat{\Delta}_-(\hat{n})=
\sqrt{\hat{n}+2}-\sqrt{\hat{n}+1}.\end{equation} By integrating
(\ref{sigma}) by using a Fresnel integral \cite{Gradshteyn}
\begin{equation}
\int_0^{\infty}dT T \sin(T^2/A) \sin(BT) =
\frac{AB}{4}\sqrt{\frac{\pi A}{2}}\left(
\cos\frac{AB^2}{4}+\sin\frac{AB^2}{4} \right),
\end{equation}
such that (with $\eta\Omega t = T$)
\begin{eqnarray}
\int_0^{\infty}dT T \sin(T^2/A) \langle  \hat{\sigma}_+ \rangle =
-\frac{i}{2}\langle\psi_F(0)|
\hat{V}^{\dagger}(\hat{\gamma}_1-\hat{\gamma}_2)|\psi_F(0)\rangle
\label{sigma2}
\end{eqnarray}
with
\begin{equation}
\hat{\gamma}_1^{(1)}  = \frac{A\hat{\Delta}_+(\hat{n})}{4}
\sqrt{\frac{\pi A}{2}}\left( \cos\left[
\frac{A\hat{\Delta}_+^2(\hat{n})}{4}\right]+\sin\left[
\frac{A\hat{\Delta}_+^2(\hat{n})}{4}\right]\right)
\end{equation}
and
\begin{equation}
\hat{\gamma}_2^{(1)} =
\frac{A\hat{\Delta}_-(\hat{n})}{4}\sqrt{\frac{\pi A}{2}}\left(
\cos\left[ \frac{A\hat{\Delta}_-^2(\hat{n})}{4}\right]+\sin\left[
\frac{A\hat{\Delta}_-^2(\hat{n})}{4}\right]\right).
\end{equation}
Now we use the approximation \cite{kurizki1,kvogel} $\sqrt{(\hat{n}+2)(\hat{n}+1)}\approx
\hat{n} +3/2$ that is valid for large photon numbers; we then can write
$\hat{\Delta}_+^2(\hat{n}) \approx 4\hat{n}+6$ and
$\hat{\Delta}_-^2(\hat{n}) \approx 0$. By setting $A=4\pi$, we obtain
\begin{equation}
\hat{\gamma}_1^{(1)} \approx
(\sqrt{\hat{n}+2}+\sqrt{\hat{n}+1})\pi \cos\left[
(4\hat{n}+6)\pi\right] = \sqrt{2}\pi^2\hat{\Delta}_+(\hat{n})
\end{equation}
and
\begin{equation}
\hat{\gamma}_2^{(1)} \approx \sqrt{2}\pi^2
\hat{\Delta}_-(\hat{n}),
\end{equation}
so that the integral transform (\ref{sigma2}) becomes
\begin{eqnarray}\label{sigmamas}
\int_0^{\infty}dT T \sin(T^2/A) \langle  \hat{\sigma}_+ \rangle &
= & \sqrt{2}\pi^2 \langle\psi_v(0)|
\hat{V}^{\dagger}\sqrt{\hat{n}+1}|\psi_v(0)\rangle
\\ \nonumber  & = & \sqrt{2}\pi^2 \langle\psi_v(0)|
\hat{a}^{\dagger}|\psi_v(0)\rangle .
\end{eqnarray}
To measure $\langle\psi_F(0)|
[\hat{a}^{\dagger}]^2|\psi_F(0)\rangle $ it is necessary a
two-phonon transition; in this case
\begin{equation}
{H}_2=\lambda^{(2)} \hat{R}^2 \left(
\begin{array}{cc}
0 & \sqrt{(\hat{n}+1)(\hat{n}+2)}
\\  \sqrt{(\hat{n}+1)(\hat{n}+2)}& 0
\end{array}
\right) [{R}^{\dagger}]^2,
\end{equation}
where $\lambda^{(2)}$ is the interaction constant in the
two-phonon case. One can find the evolution operator that will be
given by an expression similar to (\ref{evol}), just changing
$\sqrt{\hat{n}+1}\rightarrow\sqrt{(\hat{n}+1)(\hat{n}+2)}$,
$\hat{V}\rightarrow\hat{V}^2$ and
$\hat{V}^{\dagger}\rightarrow[\hat{V}^{\dagger}]^2$. It is then
easy to calculate the average of $\hat{\sigma}_+^{(2)}$,
 which is given by
\begin{eqnarray}\label{sigma2p}
\langle  \hat{\sigma}_+^{(2)} \rangle &=& -i\langle
\psi_F(0)|\cos\left[\lambda^{(2)}
t\sqrt{(\hat{n}+1)(\hat{n}+2)}\right]
[\hat{V}^{\dagger}]^2\sin\left[\lambda^{(2)}
t\sqrt{(\hat{n}+1)(\hat{n}+2)}\right]|\psi_F(0)\rangle\nonumber
\\  &=& -\frac{i}{2}\langle \psi_F(0)|[\hat{V}^{\dagger}]^2\left(\sin[\lambda
t\hat{\delta}_+(\hat{n})] - \sin[\lambda
t\hat{\delta}_-(\hat{n})]\right)|\psi_F(0)\rangle
\end{eqnarray}
with
\begin{equation}
\hat{\delta}_+(\hat{n})=
\sqrt{(\hat{n}+4)(\hat{n}+3)}+\sqrt{(\hat{n}+2)(\hat{n}+1)}\approx
2\hat{n}+5,
\end{equation}
and
\begin{equation}
\hat{\delta}_-(\hat{n})=
\sqrt{(\hat{n}+4)(\hat{n}+3)}-\sqrt{(\hat{n}+2)(\hat{n}+1)}\approx 2.
\end{equation}
Again by (Fresnel) integration of the above expression
\begin{eqnarray}\label{sigmamas}
\int_0^{\infty}dT T \sin(T^2/A) \langle  \hat{\sigma}_+
^{(2)}\rangle  =  -i4\pi^2 \langle\psi_F(0)|
[\hat{V}^{\dagger}]^2(\hat{\gamma}_1^{(2)}-\hat{\gamma}_2^{(2)})|\psi_F(0)\rangle
\end{eqnarray}
with
\begin{equation}
\hat{\gamma}_1^{(2)}  = \frac{A\hat{\delta}_+(\hat{n})}{4}
\sqrt{\frac{\pi A}{2}}\left( \cos\left[
\frac{A\hat{\delta}_+^2(\hat{n})}{4}\right]+\sin\left[
\frac{A\hat{\delta}_+^2(\hat{n})}{4}\right]\right)
\end{equation}
and
\begin{equation}
\hat{\gamma}_2^{(2)} =
\frac{A\hat{\delta}_-(\hat{n})}{4}\sqrt{\frac{\pi A}{2}}\left(
\cos\left[ \frac{A\hat{\delta}_-^2(\hat{n})}{4}\right]+\sin\left[
\frac{A\hat{\delta}_-^2(\hat{n})}{4}\right]\right).
\end{equation}
By choosing the value  $A=8\pi$, we obtain
\begin{eqnarray}\label{sigma2mas}
\int_0^{\infty}dT T \sin(T^2/8\pi) \langle  \hat{\sigma}_+
^{(2)}\rangle & = &  -i8\pi^4 \langle\psi_F(0)|
[\hat{V}^{\dagger}]^2\sqrt{(\hat{n}+1)(\hat{n}+1)}|\psi_F(0)\rangle
\nonumber \\& =&  -i8\pi^4\langle [\hat{a}^{\dagger}]^2\rangle.
\end{eqnarray}
Therefore, squeezing may be measured via this scheme.

\section{Ion-laser interaction in a trap with time-dependent frequency}
In this section, we study the problem of an ion trapped with a frequency that depends on time and interacting with a laser beam \cite{Manko2, Schrade}. Using unitary transformations, we show that this system is equivalent to a system formed by a two levels subsystem with time dependent parameters interacting with a quantized field. The procedure to build the Hamiltonian for this case is exactly the same that in the time independent frequency case (Section 3), but we have to keep in mind that now the frequency is time dependent. \\
The Hamiltonian is
\begin{equation}\label{801080}
    H=\frac{1}{2} \left[p^2 + \nu^2(t) x^2 \right] + \frac{1}{2} \omega_{21} \sigma_z +  \lambda\left[ E^{(-)(x,t)} \sigma_- + \textrm{H.C.}\right],
\end{equation}
and then the Schr\"{o}dinger equation can be written as
\begin{equation}\label{801090}
    i \frac{\partial}{\partial t} |\xi (t)\rangle =  H |\xi (t)\rangle.
\end{equation}
To solve the problem, we make the transformation
\begin{equation}\label{801100}
    |\phi(t)\rangle=T_{\textrm{SD}}(t)|\xi (t)\rangle,
\end{equation}
where
\begin{equation}\label{801110}
    T_{\textrm{SD}}(t)=\exp{\left\{\frac{i\ln\left[ \rho(t) \sqrt{\nu_0}\right](xp+px)}{2} \right\}}
    \exp{\left[ \frac{-i \dot{\rho}(t)x^2}{2 \rho(t)}              \right]},
\end{equation}
and we have also the Ermakov equation
\begin{equation}\label{801115}
\frac{d^2\rho}{dt^2}+\nu^2(t)\rho = \frac{1}{\rho^3},
\end{equation}
as an auxiliary equation. We apply the unitary transformation (\ref{801110}) to
the Hamiltonian (\ref{801080}); i.e., we must calculate the
expression
\begin{equation}\label{801120}
    H_{\textrm{SD}}= i \frac{\partial T_{\textrm{DS}}(t)}{\partial t}{T_{\textrm{SD}}}^\dagger (t) + T_{\textrm{SD}}(t) H {T_{\textrm{SD}}}^\dagger (t).
\end{equation}
We find,
\begin{equation}\begin{split}\label{801130}
    H_{\textrm{SD}}&= \frac{1}{2 \nu_0 \rho^2(t)}(p^2+\nu_0^2 x^2)+ \frac{1}{2} \omega_{21} \sigma_z \\
    & +  \Omega \left\{ \exp{\left[-i\left( k \rho(t)\sqrt{\nu_0}x^2-\omega t \right)  \right]\sigma_- + \textrm{H.C.}} \right\},
\end{split}\end{equation}
with  $  \Omega=\lambda E_0$, the Rabi frequency. \\
Using the Ermakov invariant, the time dependence of the trap has
been factorized; the time dependence is implicit in $\rho (t)$. We
go now to a frame rotating at frequency $\omega$, by means of the
unitary transformation
\begin{equation}\label{80115}
    T_\omega(t)= \exp {\left( \frac{i}{2} \omega t \sigma_z \right)}.
\end{equation}
The Hamiltonian is transformed to
\begin{equation}\begin{split}\label{801160}
    H_{\omega} & = \frac{1}{2 \nu_0 \rho^2(t)}(p^2+\nu_0^2 x^2)+ \frac{1}{2} (\omega_{21} - \omega) \sigma_z   \\
    & + \Omega(t) \left\{ \exp \left[ -i (\hat{a} + \hat{a}^\dagger) \eta(t) \right] \sigma_-   + \textrm{H.C.} \right\}.
\end{split}\end{equation}
Denoting the detuning frequency between the laser and the ion by
$\delta=\omega_{21}-\omega$, and the characteristic frequency of
the time dependent harmonic oscillator by
\begin{equation}\label{801180}
    \tilde{\omega}(t)=\frac{1}{\rho^2(t)},
\end{equation}
we get
\begin{equation}\label{801170}
    H_{\omega}= \tilde{\omega}(t)(\hat{n}+1/2)+ \frac{\delta}{2}  \sigma_z
        + \Omega(t) \left\{ \exp{\left[-i (\hat{a}+\hat{a} ^\dagger)\eta(t) \right]\sigma_- + \textrm{H.C.}} \right\}.
\end{equation}
The time dependent Lamb-Dicke parameter is
\begin{equation}\label{801190}
    \eta(t)=\eta_0\rho(t) \sqrt{\nu_0},
\end{equation}
with
\begin{equation}\label{801200}
    \eta_0=k \sqrt{\frac{1}{2\nu_0}},
\end{equation}
where $k$ is the wave vector of the laser beam. Comparing with the Hamiltonian (\ref{80420}),
 the Hamiltonian (\ref{801170}) is equivalent, but with all the parameters depending on time.

\subsection{Exact linearization of the system}
We call linearization of the system the process to reduce the
exponent of the ladder operators $\hat{a}$ and $\hat{a} ^\dagger$
to the first power, without using approximations. To this end, we
make the transformation
\begin{equation}\label{801210}
    |\phi_R\rangle = R(t) |\phi_\omega\rangle,
\end{equation}
where $R(t)$ is given by
\begin{equation}\label{801220}
    R(t)=\exp{\left[ -\frac{\pi}{4}(\sigma_+ - \sigma_-)\right]}\exp{\left[ -i\frac{\eta(t)}{2}(\hat{a}+ \hat{a} ^\dagger)\sigma_z\right]}.
\end{equation}
Transforming the Hamiltonian, we get
 \begin{equation}\label{801230}
    H_R= \nu_0 \hat{n} + \Omega \sigma_z + \left\{ \frac{\delta}{2}+i \left[ \hat{a}\beta(t)-\hat{a} ^\dagger \beta^\ast (t)\right] \right\}(\sigma_+ + \sigma_-)
 \end{equation}
 with
 \begin{equation}\label{801235}
    \beta(t)=\frac{\eta(t)\tilde{\omega}}{2}-\frac{i\dot{\eta}(t)}{2}.
 \end{equation}
 The term $\tilde{\omega}(t)/2$ has not been considered, because it is only a phase, and when the observable mean values are taken, it disappears. \\
 With the transformation (\ref{801220}) we have achieved our goal: linearize the Hamiltonian without any type of approximation. The rotating wave approximation is not used, and this leaves open the possibility to consider different intensity regimes. No assumption has been made about the Lamb-Dicke parameter $\eta(t)$. It is also valid for any type of detuning and for any time dependence of the frequency of the trap.  \\
 It is also important to remark that we have not imposed any condition in the time dependence of the frequency of the trap; in principle, this frequency can assume any temporal form. For a Paul trap, the more general form is
 \begin{equation}\label{801240}
    \nu^2(t)=a-2q\cos2t,
 \end{equation}
 and the Hamiltonian (\ref{801230}) is the ion-laser interaction  with micromotion included. Also, this Hamiltonian gives us the freedom to consider arbitrary time dependent frequencies. For instance, if we consider a sudden change in the trap frequency, we would generate squeezed states for the vibrational wave function.

\subsection{Squeezed states by changing the trap's frequency}
If we consider no interaction with a laser; i.e., $\Omega=0$, the
Hamiltonian for the ion with an arbitrary (trap) frequency is simply \cite{manko3}
\begin{equation}
H=\frac{1}{2} \left[p^2+\omega^2(t)x^2\right]
\end{equation}
and we have seen that the transformation (\ref{801110}) produces the Hamiltonian
\begin{figure}[h!]\label{Figure13}
  \centering
  \includegraphics[width=.95\textwidth,height=.5\textwidth]{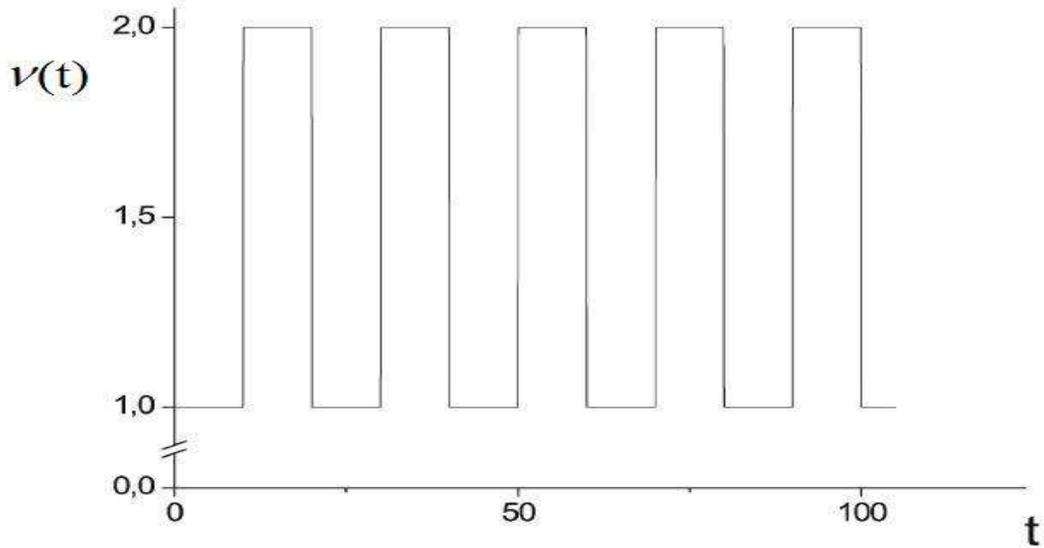}
  \caption {Trap's time dependent frequency as a train of steps.}
\end{figure}
\begin{equation}
H=\frac{1}{2\rho^2(t)}\left(p^2+x^2\right)
\end{equation}
which is in an integrable form. If we consider the vibrational motion state to
be in a coherent state, $|\alpha\rangle$, then the evolved wave function reads
\begin{equation}
|\psi(t)\rangle=e^{-i\hat{I}\int_0^t\tilde{\omega}(\tau)d\tau}T^{\dagger}_{SD}(t)T_{SD}(0)|\alpha\rangle
\end{equation}
where
\begin{equation}
\hat{I}=\frac{1}{2}\left[\frac{x^2}{\rho^2}+(\rho p-\dot{\rho}x)^2\right]
\end{equation}
is the so-called Lewis-Ermakov invariant. If we consider $\nu(t)$ to have the form as in Figure 13, we can solve numerically the Ermakov equation, and obtain the auxiliary function, $\rho$, which we plot in Figure 14; note that we will have a series of maximums and minimums. Because the transformation $T^{\dagger}_{SD}(t)$ depends on $\rho$, we can analyze from this figure the form in which the vibrational wave function is modified.  Note from (\ref{801110}) that the second factor involves the derivative of $\rho$, which for the maximums and minimums is zero, making this exponential equal to one. Therefore transformation (\ref{801110}) may be written for such times as
\begin{figure}[h!]\label{tren}
  \centering
  \includegraphics[width=.95\textwidth,height=.6\textwidth]{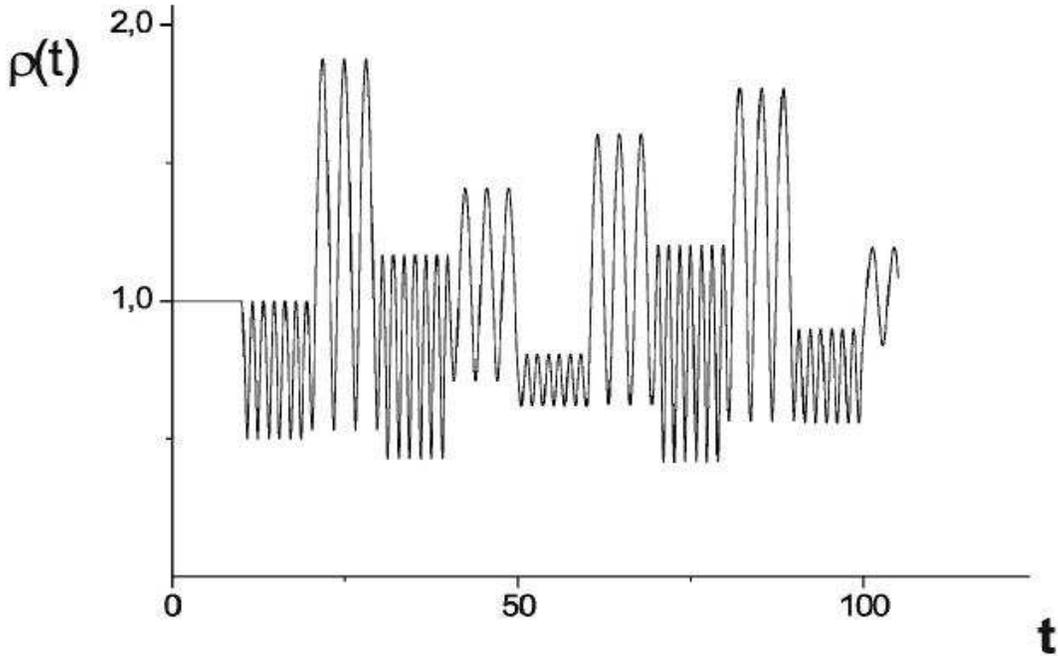}
  \caption{ Solution for the Ermakov equation for a time dependent frequency as shown in Figure 13. }
\end{figure}
\begin{equation}\label{801135}
    T_{\textrm{SD}}(t_M)=\exp{\left\{\frac{i\ln\left[ \rho(t_M) \sqrt{\nu_0}\right](xp+px)}{2} \right\}};
\end{equation}
i.e., a squeeze operator. Because $T_{SD}(0)=1$ for this form of the time dependent frequency, if initially we start with a coherent state, squeezed states will be produced at times $t_M$.

\section{Nonlinear coherent states and their modeling in photonic lattices}
Nonlinear coherent states \cite{manko,mikha} may be generated in
ion traps \cite{vogeliv} and in this way   realization of a
quantum-mechanical counterpart of nonlinear optics has been
achieved. For an appropriate laser-beam propagation geometry which
affects only the dynamics in one vibrational mode of frequency
$\nu$, in the rotating-wave approximation, the Hamiltonian
describing the effect of the Raman laser drive on the dynamics of
the vibrational mode is given by \cite{Wall97}
\begin{equation}\label{801136}
   H=\frac{\Omega}{2}\hat{g}_k(\hat{n})(i\eta a)^k+ \textrm{H.C.},
\end{equation}
with $\langle n|\hat{g}_k(\hat{n})
|n\rangle=\frac{n!}{(n+k)!}L_n^{(k)}(\eta^2)e^{-\eta^2/2}$. If we
consider the initial vibrational wave function to be in the vacuum
state, after application of the evolution operator to this initial
condition, a nonlinear coherent state is obtained. If we write the
solution for the above Hamiltonian in the form
\begin{equation}\label{801135}
   |\psi(t)\rangle= \sum_{n=0}^{\infty} u_n(t)  |n\rangle,
\end{equation}
and insert it into the Schr\"odinger equation, we obtain the following
semi-infinite system of differential equations for the amplitudes $u_n(t)$ (for $k=1$)
\begin{equation}\label{801137}
i\frac{du_0(t)}{dt}=\frac{\Omega}{2}{f}_{1}(1)u_{1}.
\end{equation}
and
\begin{equation}\label{801138}
i\frac{du_n(t)}{dt}=\frac{\Omega}{2}({g}_1(n)\sqrt{n}u_{n-1}+{g}_{1}(n+1)\sqrt{n+1}u_{n+1}), \qquad n>0
\end{equation}
\begin{figure}[h!]
  \centering
  \includegraphics[width=10cm]{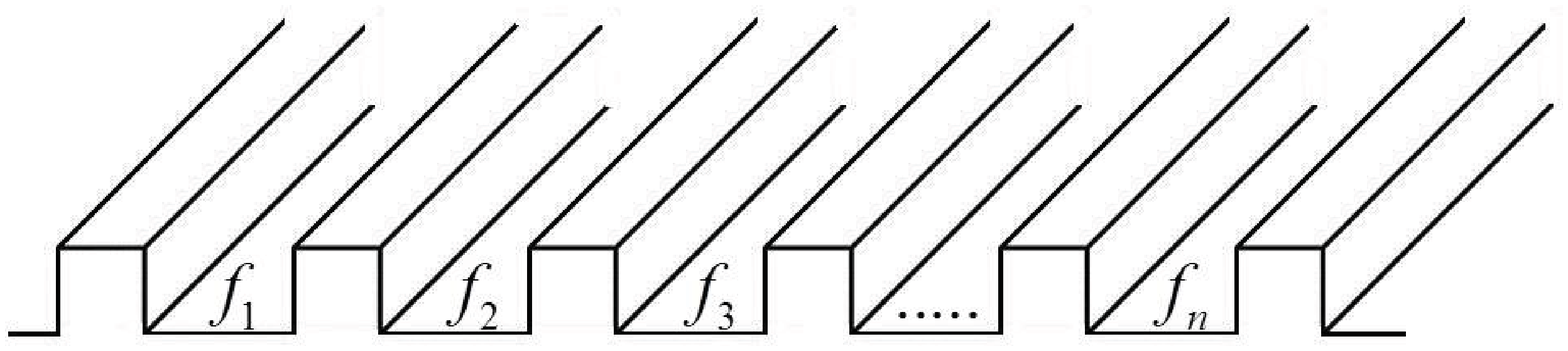}
  \vspace{-2.5cm}
  \caption{A semi-infinite waveguide array where $f_k=g_1(k)\sqrt{k}$ is the corresponding coupling constant. }
  \end{figure}
This system may be modeled in classical optics by light propagation in optical lattices as
evanescently coupled waveguides  have emerged  as a promising candidate for the realization of an ideal,
one-dimensional lattice with tunable hopping \cite{leija1,leija2}. The system of differential equations
given in (\ref{801137}) and (\ref{801138}) may be produced in a setup of waveguide arrays as the one in figure 15,
therefore modeling the ion laser Hamiltonian (\ref{801136}).

\section{Conclusions}
We have studied the ion-laser interaction, producing solutions
that do not require too many approximations and allowing  to reach
ranges of parameters, such as Lamb-Dicke, detuning and intensity,
not reached before. Exact linearization in the case of time
dependent frequency has been shown by using Ermakov-Lewis
invariant methods. In the case in which the frequency is
considered constant, we have shown that it is possible to find
exact eigenstates, access  blue and red sidebands with (a) in the
low intensity regime without considering the rotating wave
approximation and (b) in the on resonant case, by adjusting the
intensity to half the trap's frequency, i.e., solely by intensity
manipulation. We have also shown how a special class of
nonclassical states, namely, $N00N$ states may be produced in
these systems and how the ion laser interactions may be mimicked
by a particular evanescent coupling of waveguides.

\end{document}